\documentclass[12pt,preprint]{aastex}

\def\today{\ifcase\month\or January\or February\or March\or April\or
May\or June\or July\or August\or September\or October\or November\or
December\fi \space\number\day, \number\year}
\newcount\nextyear
\nextyear=\number\year \advance\nextyear by 1

\def\kms{\ifmmode{\rm km\thinspace s^{-1}}\else km\thinspace s$^{-1}$\fi}

\shortauthors{Torres et al.}
\shorttitle{BD$+05\arcdeg$706}

\begin{document}

\title{Optical photometry and X-ray monitoring of the ``Cool Algol"
BD+05$^\circ$706: Determination of the physical properties}

\author{Guillermo Torres\altaffilmark{1},
	Jeff A.\ Mader\altaffilmark{2},
	Laurence A.\ Marschall\altaffilmark{3},
	Ralph Neuh\"auser\altaffilmark{4,5},
	Alaine S.\ Duffy\altaffilmark{3,1}
}

\altaffiltext{1}{Harvard-Smithsonian Center for Astrophysics, 60 Garden
St., Cambridge, MA 02138}

\altaffiltext{2}{McDonald Observatory, Hobby-Eberly Telescope, P.O.\ Box
1337, State Highway 118 North, Ft.\ Davis, TX 79734}

\altaffiltext{3}{Department of Physics, Gettysburg College, 300 North
Washington Street, Gettysburg, PA 17325}

\altaffiltext{4}{Max-Planck-Institut f\"ur extraterrestrische Physik,
D-85740 Garching, Germany}

\altaffiltext{5}{Astrophysikalisches Institut und
Universit\"atssternwarte, Universit\"at Jena, Schillerg\"asschen 2,
D-07445 Jena, Germany}

\email{gtorres@cfa.harvard.edu}
\email{*** To appear in The Astronomical Journal, June 2003 ***}

\begin{abstract}

We present new photometric observations in the $BV\!RI$ bands of the
double-lined eclipsing binary BD$+05\arcdeg$706 conducted over three
observing seasons, as well as new X-ray observations obtained with the
ROSAT satellite covering a full orbital cycle ($P = 18.9$~days). A
detailed light-curve analysis of the optical data shows the system to
be semidetached, confirming indications from an earlier analysis by
\cite{Torres1998}, with the less massive and cooler star filling its
Roche lobe.  The system is a member of the rare class of cool Algol
systems, which are different from the ``classical" Algol systems in
that the mass-gaining component is also a late-type star rather than a
B- or A-type star. By combining the new photometry with a reanalysis
of the spectroscopic observations reported by \cite{Torres1998} we
derive accurate absolute masses for the components of $M_1 = 2.633 \pm
0.028$~M$_{\sun}$ and $M_2 = 0.5412 \pm 0.0093$~M$_{\sun}$, radii of
$R_1 = 7.55 \pm 0.20$~R$_{\sun}$ and $R_2 = 11.02 \pm
0.21$~R$_{\sun}$, as well as effective temperatures of $5000 \pm
100$~K and $4640 \pm 150$~K for the primary and secondary,
respectively. There are obvious signs of activity (spottedness) in the
optical light curve of the binary. Our X-ray light curve clearly shows
the primary eclipse but not the secondary eclipse, suggesting that the
primary star is the dominant source of the activity in the system. The
depth and duration of the eclipse allow us to infer some of the
properties of the X-ray emitting region around that star. 

\end{abstract}


\section{Introduction}
\label{sec:intro}

The class of interacting binaries known as the ``classical Algols" are
semi-detached systems that contain a late-type giant or subgiant that
fills its Roche lobe and is transferring mass onto a much more
luminous and more massive B--A main sequence star.  A new class of
binaries has been found with properties that are similar to those of
the classical Algols, except that the mass gaining component is a
late-type giant or subgiant like its companion.  These systems were
first recognized by \citet{Popper1992} and are now referred to as the
``cool Algols." They typically display signs of magnetic activity in
the form of Ca II H$+$K emission, H$\alpha$ emission, star spots, and
strong X-ray emission.  Only about a dozen cool Algols are known, and
many of them were mistakenly assigned to the larger group of RS~CVn
binaries which contain a luminous K-type primary with a fainter F--G
type companion.  The main difference between the cool Algols and the
RS~CVn systems is that the former are undergoing Roche lobe overflow,
while the latter are detached \citep[see,
e.g.,][]{Popper1980,Hall1989}. 

BD$+05\arcdeg$706 (RXJ0441.9$+$0537, PPM~147827, $\alpha = 4^{\rm h}
41^{\rm m} 57\fs64$, $\delta = +5\arcdeg 36\arcmin 34\farcs3$, J2000,
$V = 9.4$) is one of the recent additions to this rare category of
cool Algols, and was discovered in the course of spectroscopic
follow-up observations of a sample of X-ray sources selected from the
ROSAT All-Sky Survey south of the Taurus-Auriga star-forming region
\citep{Neuhauser1997}.  Although that X-ray sample was originally
designed to favor the detection of T~Tauri stars, which are also quite
active, BD$+05\arcdeg$706 was quickly found to be of an entirely
different nature. \citet{Torres1998} (hereafter Paper~I) reported a
double-lined spectroscopic orbital solution with a period of
18.9~days, a very small and possibly spurious eccentricity, and a mass
ratio of $q = 0.21$ ($M_{\rm secondary}/M_{\rm primary}$), typical of
the Algol systems. 

The system contains two cool giants. The \ion{G8}{3} ``primary" (the
more massive star) is detached, while the \ion{K1-K2}{3} ``secondary"
has filled its critical lobe or may overflow it (Paper~I).  The star
we have called the ``secondary" is currently the larger, less massive,
and cooler component, as in the classical Algols.  Spectroscopic
convention is used throughout this paper instead of the notation used
in papers on stellar evolution, which refer to the initially more
massive star as the primary.  The general properties of the system
were described in Paper I. Although photometry for BD$+05\arcdeg$706
was not available at the time of that study, eclipses were estimated
to be very likely.  Subsequent photometric monitoring in the optical
confirmed this prediction \citep{Marschall1998}. 

The general characteristics of the cool Algols are fairly well defined
\citep[see][and Paper~I]{Popper1992}, but accurate information on
individual systems is still far from complete.  Only five of the
members, aside from BD$+05\arcdeg$706, have spectroscopic orbits for
both components \citep[RZ~Cnc, AR~Mon, RT~Lac, AD~Cap, and
RV~Lib;][]{Popper1976,Popper1991}, and light curve analyses for the
first three of these have yielded the absolute masses and radii
necessary in order to understand their properties.  The remaining five
objects identified as possible members of this class \citep[UZ~Cnc,
V1061~Cyg, AV~Del, GU~Her, and V756~Sco;][]{Popper1996} have very
little in the way of observations and in some cases even the orbital
periods are poorly known. 

The high-quality spectroscopic orbit presented for BD$+05\arcdeg$706
in Paper~I, combined with the fact that the system is eclipsing,
offers the opportunity to determine the absolute dimensions of the
components very accurately. To that end, photometric observations in
the $BV\!RI$ system were continued after the initial report by
\citet{Marschall1998}. We present here a full analysis of those data,
and also a re-analysis of the spectroscopic material in Paper~I with
improved techniques and the benefit of the additional information
provided by the light-curves. The latter offer a means for evaluating
subtle distortions in the Doppler measurements due to the close
proximity of the stars. As a result, we are able to determine here the
absolute masses of both components with uncertainties smaller than
2\%, and the radii with errors smaller than 3\%, which are now the
best determinations for any system of the cool Algol class. 

The global X-ray properties and the activity level of
BD$+05\arcdeg$706 (\ion{Ca}{2} H and K emission, variable H$\alpha$
line strength) were studied in some detail in Paper~I, where it was
established that the hotter primary star is most likely the site of
the bulk of the activity in the binary.  Additional X-ray observations
for the system were secured by us with the ROSAT satellite essentially
over a complete 18.9-day cycle (yielding an X-ray light curve), and
clearly show the primary eclipse but not the secondary eclipse. An
analysis of these data is presented here as well, allowing us to
obtain information about the geometry of the X-ray emitting region
surrounding the active star. 
 
\section{Observational material}
\label{sec:obs}

The photometric observations of BD$+05\arcdeg$706 in the optical were
conducted at the Gettysburg College Observatory (Gettysburg, PA) with
a 16-inch f/11 Ealing Cassegrain reflector. The detector was a
Photometrics thermoelectrically-cooled CCD camera with a
front-illuminated Thompson~7896 chip with $1024\times 1024$ pixels.
BD$+05\arcdeg$706 was observed on 94 nights during the 1997--1999
observing seasons.  A total of 80, 93, 133 and 137 observations were
made in the Johnson $B$, $V$, $R$, and $I$ filters, respectively.  The
comparison and check stars used are TYC~95-1495-1 (GSC~00095-1495) and
HD~29764 (GSC~00091-00262, BD$+05\arcdeg$704, SAO~111980),
respectively.  Basic properties for these stars are listed in
Table~\ref{tab:compcheck}. Coverage of the primary eclipse in the $B$
and $V$ filters was very poor during the first two seasons (before JD
2,451,000), and therefore those data were not used in the analysis
below. 

The precision of an individual differential photometric measurement is
estimated to be 0.010~mag, 0.010~mag, 0.009~mag, and 0.010~mag in the
$B$, $V$, $R$, and $I$ passbands, respectively, derived from the
comparison and check stars under the assumption that they do not vary.
The observations in the sense $\langle$variable minus
comparison$\rangle$ are listed in Table~\ref{tab:obsb},
Table~\ref{tab:obsv}, Table~\ref{tab:obsr}, and Table~\ref{tab:obsi}.
The depth of the eclipses in the $V$ band is approximately 0.52~mag
for the primary minimum and 0.34~mag for the secondary.  Changes in
the overall brightness at the quadratures are fairly obvious from
season to season. They are not unexpected given the activity level
displayed by the system, and will be discussed in more detail in
\S\ref{sec:phot}. 

The spectroscopic material used here is the same as that reported in
Paper~I. Briefly, it consists of 41 single-order echelle spectra
obtained from 1994 March to 1996 March with the 1.5-m Wyeth reflector
at the Oak Ridge Observatory (Harvard, Massachusetts), the 1.5-m
Tillinghast reflector at the Fred L.\ Whipple Observatory (Mount
Hopkins, Arizona), and the Multiple Mirror Telescope (also atop Mount
Hopkins, Arizona) prior to its conversion to a monolithic 6.5-m
mirror. The spectra cover 45~\AA\ with a central wavelength of
5187~\AA, and the resolving power is $\lambda/\Delta\lambda \sim
35,\!000$. The reductions to obtain radial velocities for both
components are slightly different from those used in Paper~I, and are
described later. 

X-ray observations of BD$+05\arcdeg$706 were obtained with the High
Resolution Imager instrument \citep[HRI;][]{David1996} aboard the
ROSAT satellite over a period of 18 days covering essentially a full
orbital cycle of the binary, from 1997 August 18 to September 5. The
measurements were made at intervals of one day, with the exception of
a 3-day gap due to scheduling constraints. The exposure times were
typically about 3.5~ksec each, achieving signal-to-noise (S/N) ratios
ranging from 6 to 15 per broad-band observation (0.1 to 2.4~keV), with
a mean S/N of 10. The observations were reduced with the Extended
Scientific Analysis Software \citep[EXSAS,][]{Zimmermann1994} version
2001 running under ESO-MIDAS version 01FEB.  In addition to
considering the full energy range of ROSAT, we subsequently divided
the recorded counts into a soft band including detector channels 0
through 3 (0.1 to 0.5 keV) and a hard band including channels 4
through 15 (0.6 to 2.4 keV) in order to obtain additional information
from the analysis of these data. We performed standard local and map
source detection in all three spectral bands.  The S/N ratios range
from 4 to 10 in the soft band and from 4 to 11 in the hard band.  This
allows us to define a ``hardness ratio" ${\it HR} = (H-S)/(H+S)$,
where $H$ and $S$ are the hard and soft count rates, respectively
\citep[see][]{Huelamo2000}. 

The individual X-ray observations are listed in Table~\ref{tab:xray}.
The columns give the heliocentric Julian date and the exposure times,
followed by the count rates, the corresponding errors, and the Maximum
Likelihood ({\it ML\/}) estimator for the soft, hard, and broad bands
\citep{Cruddace1988}.  The quantity {\it ML\/} provides a measure of
the existence of the source above the local background. For example, a
Maximum Likelihood of 7.4 (14.3) corresponds to a 3.5$\sigma$
(5$\sigma$) detection. The last three columns give the hardness ratio
and its corresponding uncertainty, as well as the phase in the orbit
of the binary\footnote{Since the soft-, hard-, and broad-band counts
are computed separately by the Maximum Likelihood detection algorithm,
the soft and hard counts in Table~\ref{tab:xray} do not exactly add up
to the broad-band counts, although the difference is minimal.  The
mean local background is estimated separately in each band and then
subtracted from the corresponding total counts, which contributes to
the slight discrepancy.} . 

\begin{figure}
\epsscale{1.1}
\plotone{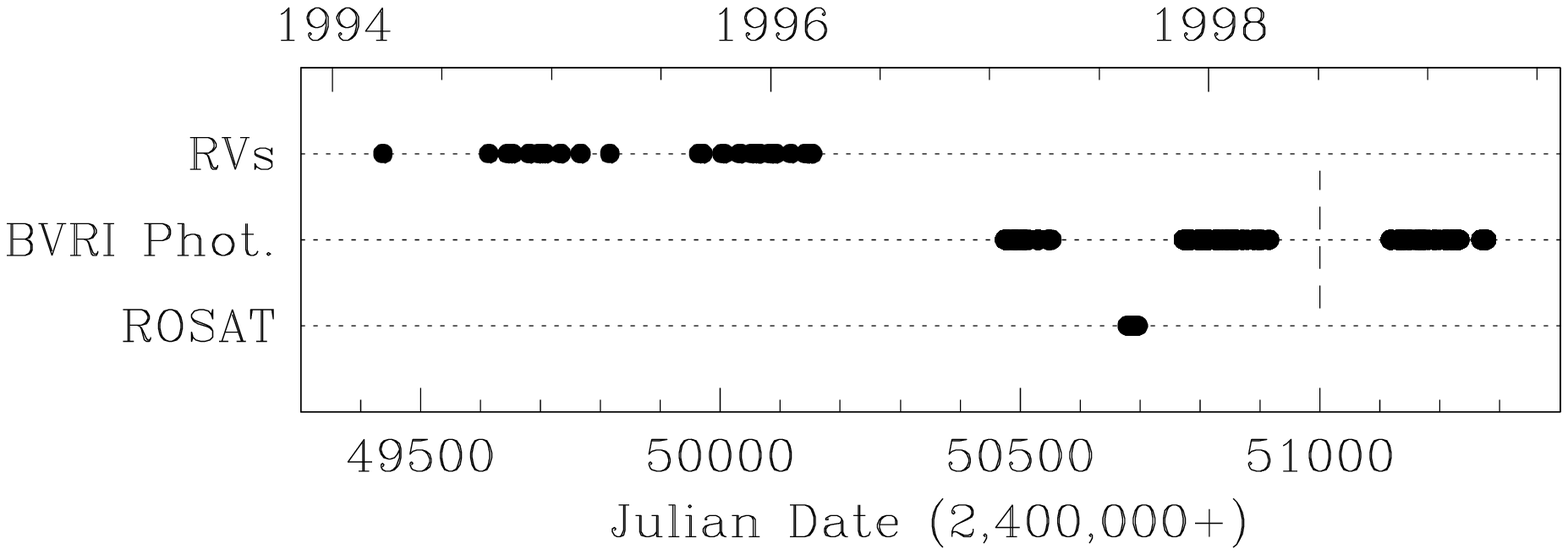}
\vskip -3.5in
 \caption[Torres.fig1.ps]{Time distribution of our spectroscopic,
photometric, and X-ray observations of BD$+05\arcdeg706$. The vertical
dashed line indicates how we have divided the optical photometry
measurements for analysis (see text).\label{fig:hist}}
 \end{figure}

Figure~\ref{fig:hist} shows the time history of the optical
photometry, the spectroscopy, and the X-ray observations.  Because of
the variability of the object, the time coverage of the optical light
curves and the radial velocities is relevant for the modeling of the
system, as we describe in \S\ref{sec:spec}. In the analyses below all
data were phased with the ephemeris
 \begin{equation}
{\rm Min~I} = {\rm HJD}~2,\!449,\!924.582~(\pm 0.012) + 18.8988~(\pm 0.0011)\times E~,
\end{equation}
 where the epoch refers to the time of primary eclipse (more massive
star occulted by the other star), and $E$ is the number of cycles
counted from this epoch. This ephemeris was determined from the radial
velocity observations described in \S\ref{sec:spec}, and supersedes the
one given in Paper~I. 

\section{Analysis of the optical photometry}
\label{sec:phot}

Examination of the raw light curves for BD$+05\arcdeg$706 revealed
significant changes over the three observing seasons that precluded
their merging into a single data set for analysis. The extent of the
changes is such that, for example, during the 1997--1998 season the
light level in the $V$ band at the first quadrature was lower than at
the second quadrature, whereas by the following season the situation
had reversed. Variations of this nature are common in active systems,
and are well explained by the presence of surface features (spots) on
one or both stars.  In order to minimize the scatter, a compromise
between the number of observations, the completeness of the phase
coverage (particularly at the minima), and the intrinsic variations
was made by dividing the observations into two data sets at Julian
Date $2,\!451,\!000$ (see Figure~\ref{fig:hist}).  As mentioned above,
the measurements in the $B$ and $V$ filters in data set \#1 do not
cover the primary eclipse, and they were therefore not used for the
final fits. Light curve analyses were carried out separately for each
bandpass in each data set ($RI$ for data set \#1, and $BV\!RI$ for
data set \#2), and also for the combined data. 

The light curve solutions were performed with the Wilson-Devinney code
\citep{WilsonDevinney1971,Wilson1979,Wilson1990} in a mode appropriate
for semi-detached systems, in accordance with the information
available for the stars from Paper~I.  Specifically, we used ``Mode~5"
for a secondary that is filling its limiting lobe.  Detached and
overcontact configurations were tried as well, but the solutions were
never as satisfactory as those in the semi-detached mode. They always
converged towards the secondary star filling its Roche lobe and the
primary well within its critical surface, in agreement with the
assessment in Paper~I.  Solutions were found for the modified
gravitational potential of the primary ($\Omega_1$, in the
Wilson-Devinney usage), the inclination angle ($i$), the relative
monochromatic luminosity of the primary ($L_1$), the mean temperature
of the secondary ($T_2$), and a phase offset ($\Delta\phi$).  The
temperature of the primary was held fixed at the spectroscopic value
of $T_1 = 5000$~K, as was the mass ratio $q\equiv M_2/M_1 = 0.2055$
(see \S\ref{sec:spec}). We have assumed the orbit to be circular,
consistent with the results of the new spectroscopic analysis below.
The monochromatic luminosity of the secondary ($L_2$) was computed by
the program directly from the temperatures of the primary and
secondary, the luminosity of the primary, the radiation laws, and the
geometry of the system.  Limb-darkening coefficients were interpolated
from the tables by \citet{vanHamme1993} for the appropriate
temperatures and for surface gravities of $\log g_1 = 3.0$ and $\log
g_2 = 2.0$ (see \S\ref{sec:spec}). A value of 0.5 was adopted for the
bolometric albedo of both components, appropriate for stars with
convective envelopes. Gravity brightening coefficients were
interpolated from the models by \citet{Claret2000}. 

Aside from the usual light elements, the distortions mentioned above
in the light curves suggested the presence of one or more spots
requiring additional free parameters. The Wilson-Devinney code is
capable of accounting for spots (assumed to be circular in shape)
based on a simple representation with four additional parameters: the
longitude ($l$, in degrees) measured counter-clockwise (as seen from
above) from the line joining the centers of the stars, the latitude
($b$, in degrees), the angular radius ($r$, in degrees), and a
temperature factor (${\it TF} = T_{\rm spot}/T_{\rm star}$)
representing the contrast in temperature between the spot and the
surrounding photosphere. 

Numerical experiments showed that in our case the fits cannot usually
distinguish the component on which the spots are located, since
similarly good solutions resulted with the spots on either star.  In
the RS~CVn systems the spots are typically located on the cooler star,
which is usually the more active component.  There are a number of
cases, however, in which the spots have instead been shown to be on
the hotter star
\citep[e.g.,][]{Hill1990,Vivekananda1991,Kjurkchieva2000,Albayrak2001}.
In BD$+05\arcdeg$706 there are strong indications that the activity is
in fact associated with the hotter star (the primary), as evidenced by
the H$\alpha$ emission (see Paper~I) and also the X-ray properties
discussed later in \S\ref{sec:xray}. It is therefore not unreasonable
to assume in the following that the spottedness in this system is
primarily associated with the hotter star.  More complicated models
with spots on both stars are possible, but are probably not justified
given the data available.  

The maximum brightness outside of eclipse is much the same for data
sets \#1 and \#2, although it occurs at phase 0.75 in the first set
and at phase 0.25 in the second. The lower brightness levels at the
other quadrature in each data set are not quite as similar to each
other and show more scatter in the photometric observations. This
suggests that the lower light levels may be due to one or more cool
spots (which tend to change with time in other similarly active
stars), as opposed to hot spots being the cause of the higher
brightness at the other quadratures.  Spots that are cooler than the
photosphere have typically been found also in virtually all RS~CVn
systems \citep[although warm features have been reported in a few
cases; see, e.g.,][]{Strassmeier2003}.  Thus we will assume here that
the spots are cool. The situation for BD$+05\arcdeg$706 obviously
calls for the cool spots to be located at roughly opposite longitudes
in the two epochs, and we found that a single spot at each epoch gave
an adequate representation of the light curves, within the
observational errors. 

Further tests showed that, due to the strong correlation between the
size and temperature of the spots, it was not possible to determine
both simultaneously from the observations at our disposal.  Grids of
preliminary solutions at fixed values of {\it TF\/} from 0.5 to 1.0
indicated a slight preference for a spot in data set \#1 with ${\it
TF} = 0.84$ (800~K cooler than the photosphere), and a similar spot in
data set \#2 with ${\it TF} = 0.87$ (650~K cooler than the
photosphere). These values are within the typical range seen in other
RS~CVn binaries \citep{ONeal1996,Schrijver2002}.  Therefore, in the
following we have held {\it TF\/} fixed at these values, and the spot
radius was left free. 

We also found that the sensitivity of the solutions to the latitude of
the spot is very weak. Experiments with fixed values of $b$ from
0\arcdeg\ to 90\arcdeg\ indicated only a very slight preference for
mid-latitude features, with fits at higher or lower latitudes being
almost indistinguishable. Although spots on the Sun are usually found
at low latitudes, Doppler imaging studies have shown that RS~CVn
systems tend to display high-latitude features and even spots that
straddle the pole \citep[e.g.,][]{Hatzes1995,Vogt1999}.  While
BD$+05\arcdeg$706 is similar to the RS~CVn binaries in many respects,
the latter systems typically rotate much more rapidly \citep[see,
e.g.,][]{Strassmeier2002}. Theoretical studies suggest that rapid
rotation in convective stars leads to the emergence of spots at higher
latitudes \citep{Schussler1992,Schussler1996,Granzer2000}, and there
is some observational support for this trend
\citep[see][]{Hatzes1998}.  Therefore, for this study we have adopted
a fixed latitude of 45\arcdeg, in agreement with the above and also
with the theoretical work by \cite{Granzer2002} that focuses
specifically on evolved stars similar to those in BD$+05\arcdeg$706. 

Initially light curve solutions were obtained separately for each
bandpass in each data set ($RI$ for data set \#1, and $BV\!RI$ for
data set \#2), to evaluate the consistency between them.  The results
were found to be reasonably similar, except that the second data set
showed a larger scatter in the temperature of the secondary ($T_2$)
and in the potential of the primary ($\Omega_1$).  Solutions were also
obtained from the combined $RI$ and $BV\!RI$ observations from data
sets \#1 and \#2, respectively, which were also quite similar to each
other.  Since the main geometric properties ($\Omega_1$ and $i$) are
not expected to change from season to season, the results for these
two parameters from the two data sets were averaged (with weights
proportional to the internal errors), and fixed in subsequent
analyses. We then repeated the fits and computed the average secondary
temperature ($T_2 = 4638$~K) from the simultaneous multi-band
analyses.  Finally, we fixed $T_2$ to this value, and we solved for
the remaining parameters including the size ($r$) and longitude ($l$)
of the spots, which clearly do change with time.  The results are
given in Table~\ref{tab:ds1} and Table~\ref{tab:ds2} for data set \#1
and \#2, respectively.  We show the separate fits for each filter
before fixing any of the parameters, and also the simultaneous
multi-band analyses in the last column calculated with $\Omega_1$,
$i$, and $T_2$ set to their average values. Also listed are the
fractional radii ($r_{\rm point}$, $r_{\rm pole}$, $r_{\rm side}$,
$r_{\rm back}$, in terms of the separation) as well as the relative
radius of a spherical star with the same volume as the distorted stars
($r_{\rm vol}$), for the primary and secondary.  The uncertainties
given in these tables are standard errors as reported by the
Wilson-Devinney code.  Experiments were also carried out to explore
the significance of third light ($L_3$), but in all cases the results
were consistent with $L_3 = 0$. 

Both eclipses in BD$+05\arcdeg$706 are partial, with approximately
65\% of the light of the primary being blocked at phase 0.0. The
latter star is nearly spherical in shape since the difference between
its polar radius and the radius directed towards the inner Lagrangian
point is less than 0.5\%. Parameters $r_{\rm point}$, $r_{\rm pole}$,
$r_{\rm side}$, and $r_{\rm back}$ in Table~\ref{tab:ds1} and
Table~\ref{tab:ds2} confirm that the secondary is distorted because it
fills its Roche lobe. 

\begin{figure}
\epsscale{0.85}
\vskip -0.5in
\plotone{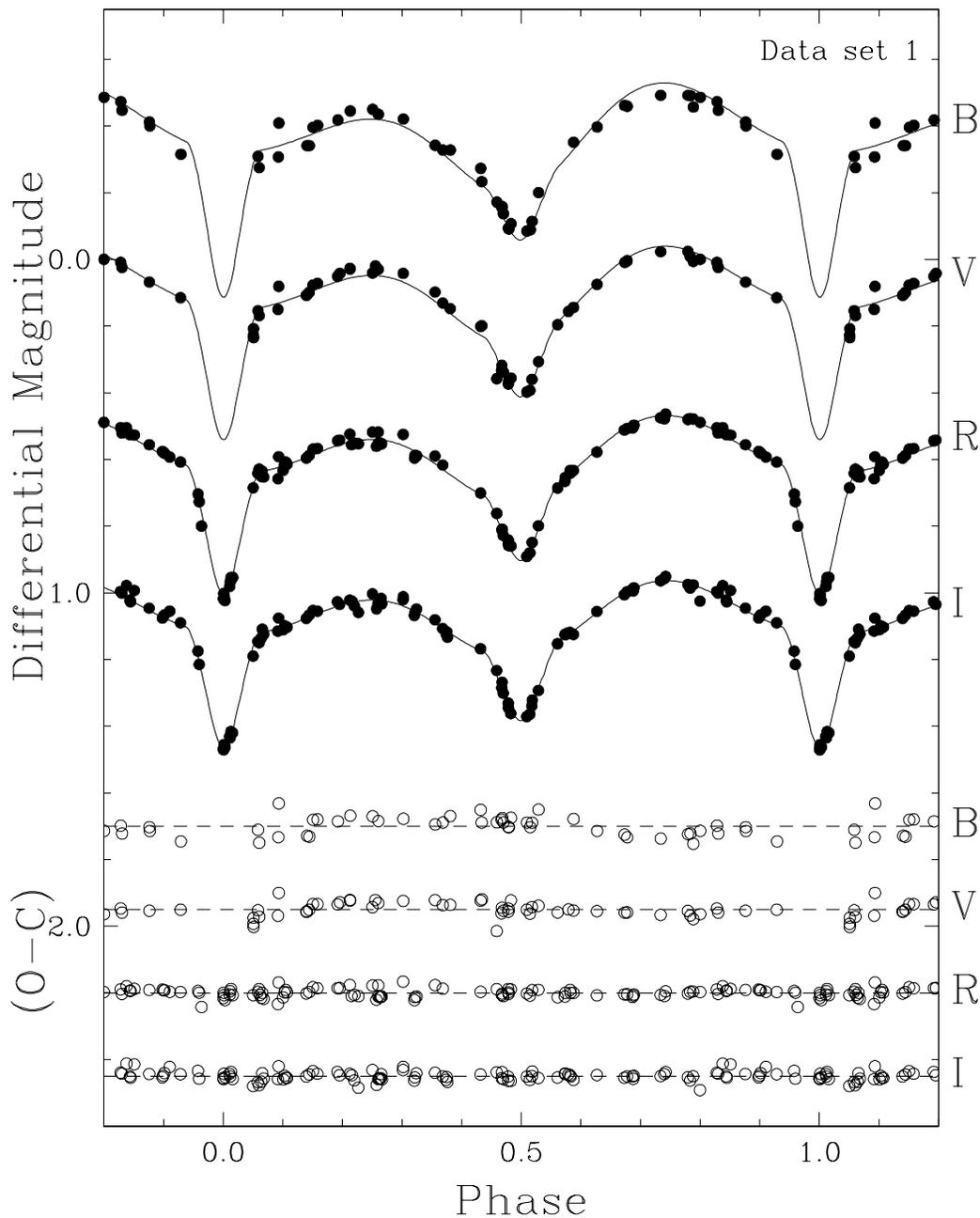}
\vskip -0.2in
 \caption[Torres.fig2.ps]{Photometric observations and light curve
solutions for data set \#1, computed from a simultaneous $RI$ fit.
The $B$ and $V$ observations, in which coverage of the primary minimum
is lacking, were not used in the final solution (see text).  The $V$,
$R$, and $I$ light curves are displaced vertically for display
purposes.  Residuals from the fits are shown at the
bottom.\label{fig:lc1}}
 \end{figure}

\begin{figure}
\epsscale{0.85}
\vskip -0.5in
\plotone{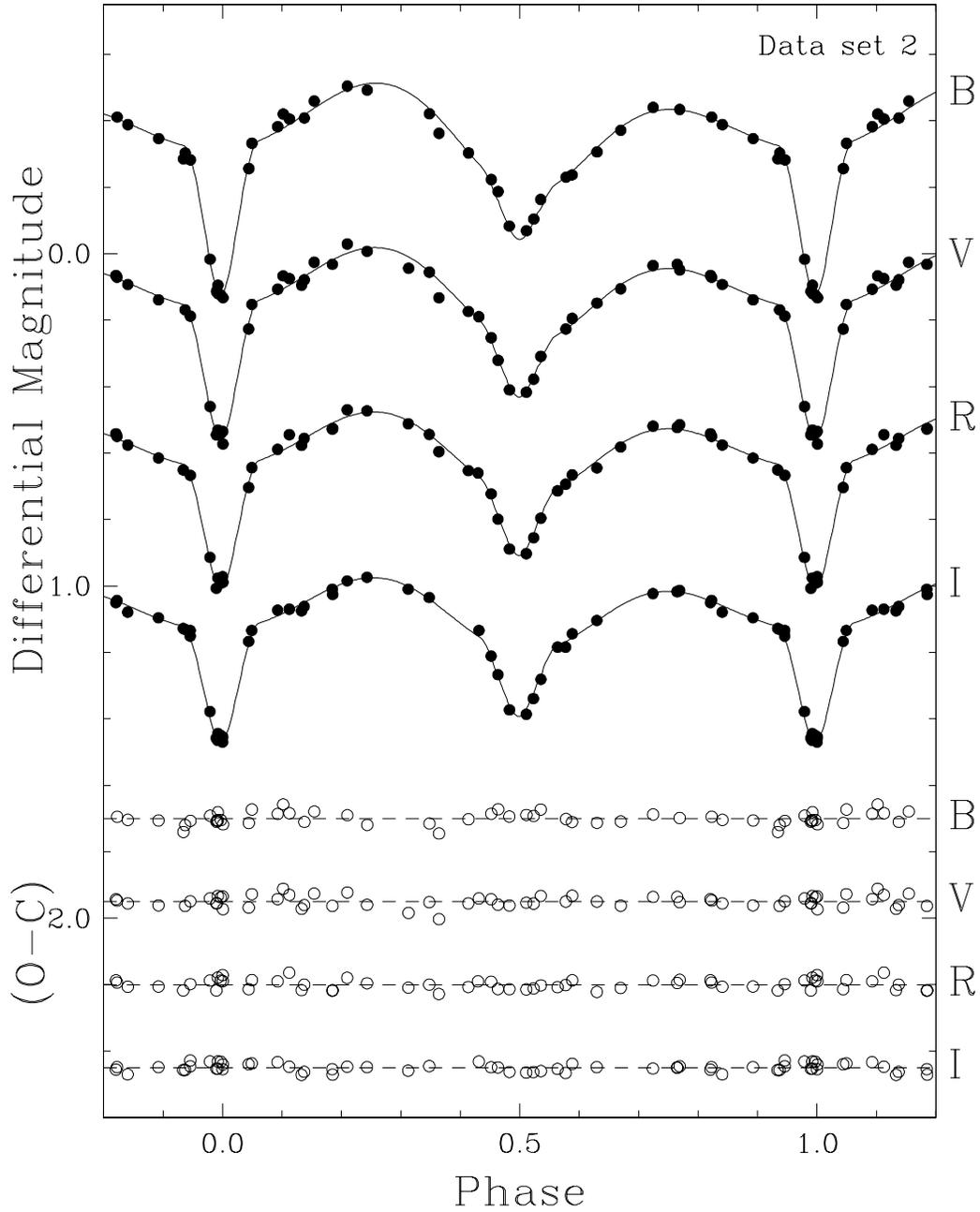}
\vskip -0.2in
 \caption[Torres.fig3.ps]{Photometric observations and light curve
solutions for data set \#2, computed from a simultaneous $BV\!RI$ fit.
The $V$, $R$, and $I$ light curves are displaced vertically for
display purposes.  Residuals are shown at the bottom.\label{fig:lc2}}
 \end{figure}

A graphical representation of the observations and light curve
solutions is given in Figure~\ref{fig:lc1} and Figure~\ref{fig:lc2}
for data sets \#1 and \#2. The $O\!-\!C$ residuals are shown at the
bottom for each passband.  The $B$ and $V$ light curves in data set
\#1 lack coverage at the primary minimum, and do not allow one to
determine any of the parameters reliably. Although we show them here
for completeness, these observations were not used in any of the fits
described above.  For these $B$ and $V$ solutions all geometric and
radiative parameters as well as the spot parameters were adopted from
the combined $RI$ solution, and only the primary luminosity was
allowed to vary. The fits are consequently poorer. 

\begin{figure}
\vskip -2in
\epsscale{1.0}
\plotone{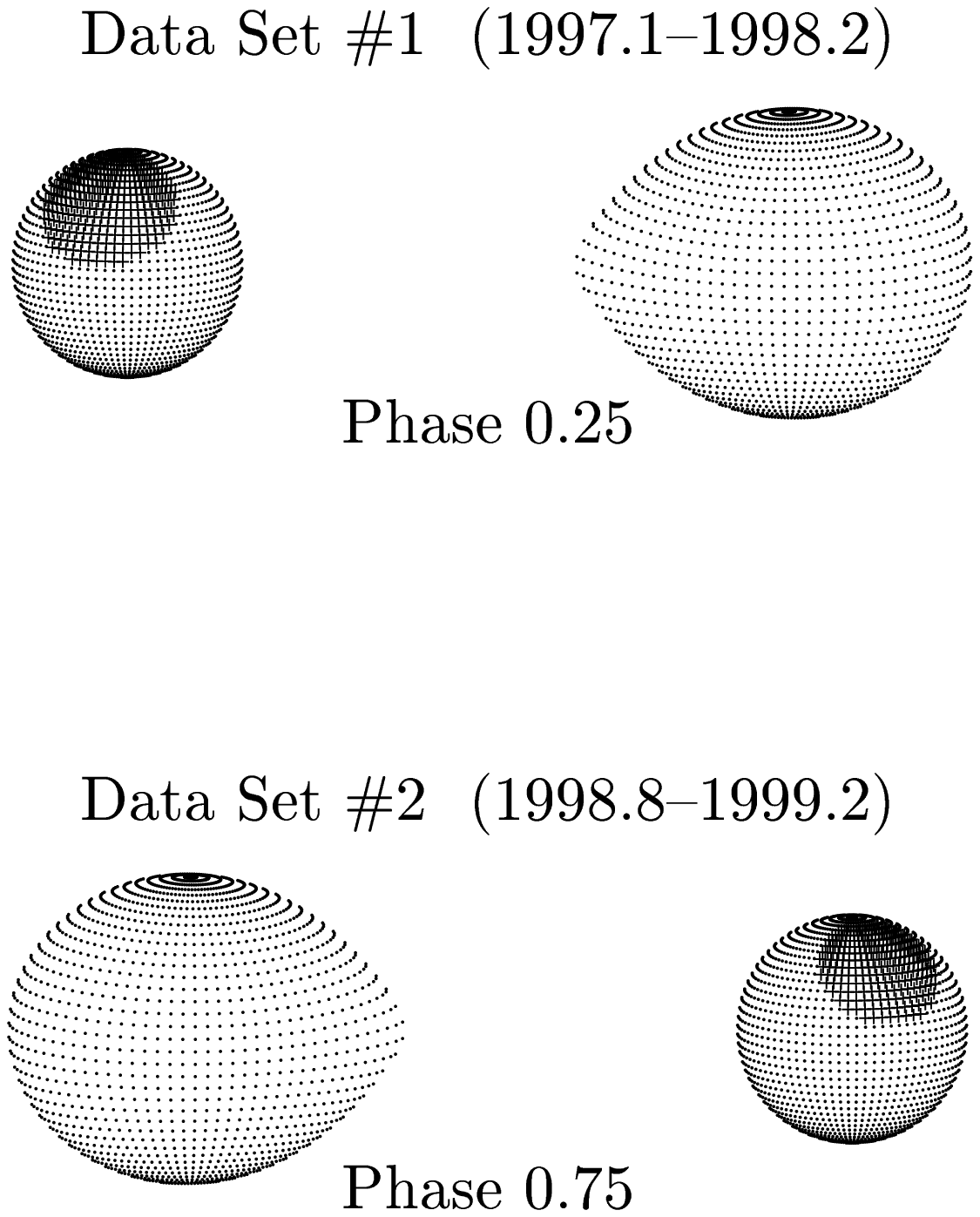}
\vskip -2.3in
 \caption[Torres.fig4.ps]{Representation of the cool spots on the
primary star at the mean epoch of data sets \#1 and \#2. The size and
separation of the stars are drawn to scale.\label{fig:spotpic}}
 \end{figure}

The spots that were found to provide a good fit to the observations in
both data sets are represented in Figure~\ref{fig:spotpic}, with the
size and separation of the stars shown to scale. The difference in
longitude between one epoch and the other (separated by approximately
1.4 years) is about 140\arcdeg.  The surface area covered by the spots
is in line with what is seen in other active systems, and reaches
$\sim$20\% of one hemisphere of the primary star in data set \#1, and
$\sim$17\% in data set \#2. 

\section{Spectroscopic analysis}
\label{sec:spec}

The procedures used here for the reduction of the spectroscopic
material are similar to those described in Paper~I in that we derived
radial velocities for the two components using TODCOR
\citep{Zucker1994}, a two-dimensional cross-correlation technique well
suited to our relatively low S/N spectra.  However, a number of
improvements were introduced that justified a re-analysis of the
original data. First, we now have at our disposal a new library of
synthetic spectra based on more recent model atmospheres by R.\ L.\
Kurucz\footnote{Available at {\tt http://cfaku5.cfa.harvard.edu}.},
which provide a somewhat better match to the spectra of real stars.
The use of these new templates has a small effect on the temperature
and rotational velocity ($v \sin i$) derived for the components of
BD$+05\arcdeg$706, and can potentially affect the radial velocities as
well. 

We re-determined the $v \sin i$ values by running extensive grids of
correlations as described in Paper~I. The results were $v_1 \sin i =
23~\kms$ and $v_2 \sin i = 31~\kms$, with estimated errors of 1~\kms\
and 2~\kms, respectively.  The primary value is 1~\kms\ larger than in
Paper~I, and the secondary value is unchanged. The new effective
temperatures, derived from similar grids, were $T_1 = 5000$~K and $T_2
= 4600$~K, with uncertainties of 100~K. These differ by less than
100~K from the determinations in Paper~I. The new primary value was
adopted for the light curve solutions described above. 

A second improvement has to do with residual systematic errors in the
raw velocities, mostly due to the relatively narrow spectral window of
these observations. Experience with similar spectroscopic material for
other binary systems has shown that these errors can be quite
significant in some cases \citep{Torres1997,Torres1999,Lacy2000}.  We
have investigated these errors for the present system by means of
numerical simulations that were discussed in detail by
\citet{Latham1996}. Briefly, we generated a set of artificial binary
spectra by combining the primary and secondary templates in the
appropriate ratio and applying velocity shifts for both components as
computed from a preliminary orbital solution at the actual times of
observation of each of our spectra.  These artificial spectra were
then processed with TODCOR in exactly the same way as the real
observations, and the resulting velocities were compared with the
input (synthetic) values.  The differences $\langle$TODCOR minus
synthetic$\rangle$ are typically under 0.5~\kms, but they are
systematic in nature. Even though the effect on the masses is
relatively small in this particular case, we have nevertheless applied
these differences as corrections to the measured velocities of
BD$+05\arcdeg$706. The corrected velocities are listed in
Table~\ref{tab:rvs} (columns 2 and 3). The differences with the values
reported in Paper~I are generally below 1~\kms. 

\begin{figure}
\epsscale{1.0}
\plotone{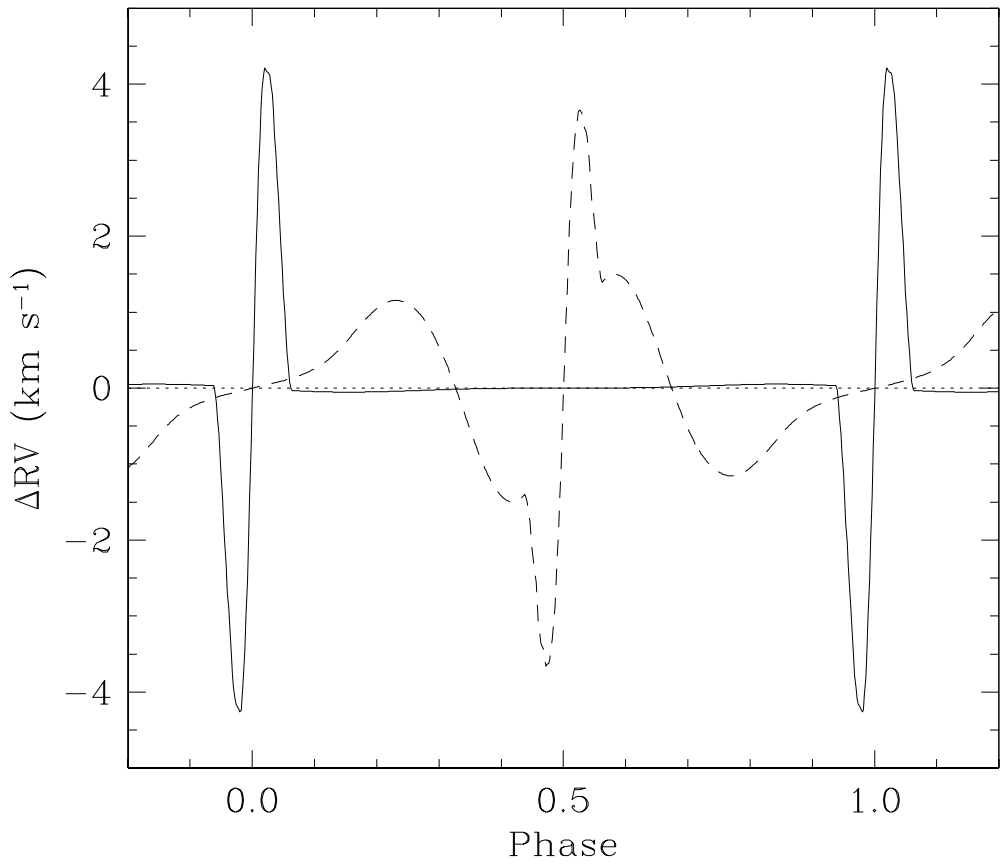}
\vskip -0.5in
 \caption[Torres.fig5.ps]{Effect of the distortions in the shape of
the stars and the Rossiter effect on the measured radial velocities of
the primary (solid curve) and secondary (dashed curve) of
BD$+05\arcdeg706$.\label{fig:rvdist}}
 \end{figure}

As a result of the proximity and distorted shapes of the stars (see,
e.g., Figure~\ref{fig:spotpic}), the radial velocities measured for
the components of BD$+05\arcdeg$706 are biased to some extent because
of the loss of symmetry in the intensity distribution across the disks
of the stars. In addition, the observations collected during the
eclipses suffer from the well-known Rossiter effect
\citep{Rossiter1924,McLaughlin1924}, which depends on the rotation
rate.  The combined magnitude of these two effects on the measured
velocities, computed with the aid of the Wilson-Devinney program, is
illustrated in Figure~\ref{fig:rvdist}.  The change in the primary
velocities is essentially confined to the eclipses (and is mostly due
to the Rossiter effect since that star is not far from the spherical
shape), but the secondary velocities show changes throughout the
entire orbital cycle because of the highly distorted shape of the
Roche-lobe filling star. The corrections for the case of
BD$+05\arcdeg$706 can reach values as large as 4~\kms, and are listed
in columns 4 and 5 of Table~\ref{tab:rvs}. The final values for the
radial velocities used below, after applying these adjustments, are
given in columns 6 and 7. 
	
\begin{figure}
\vskip -1in
\epsscale{1.0}
\plotone{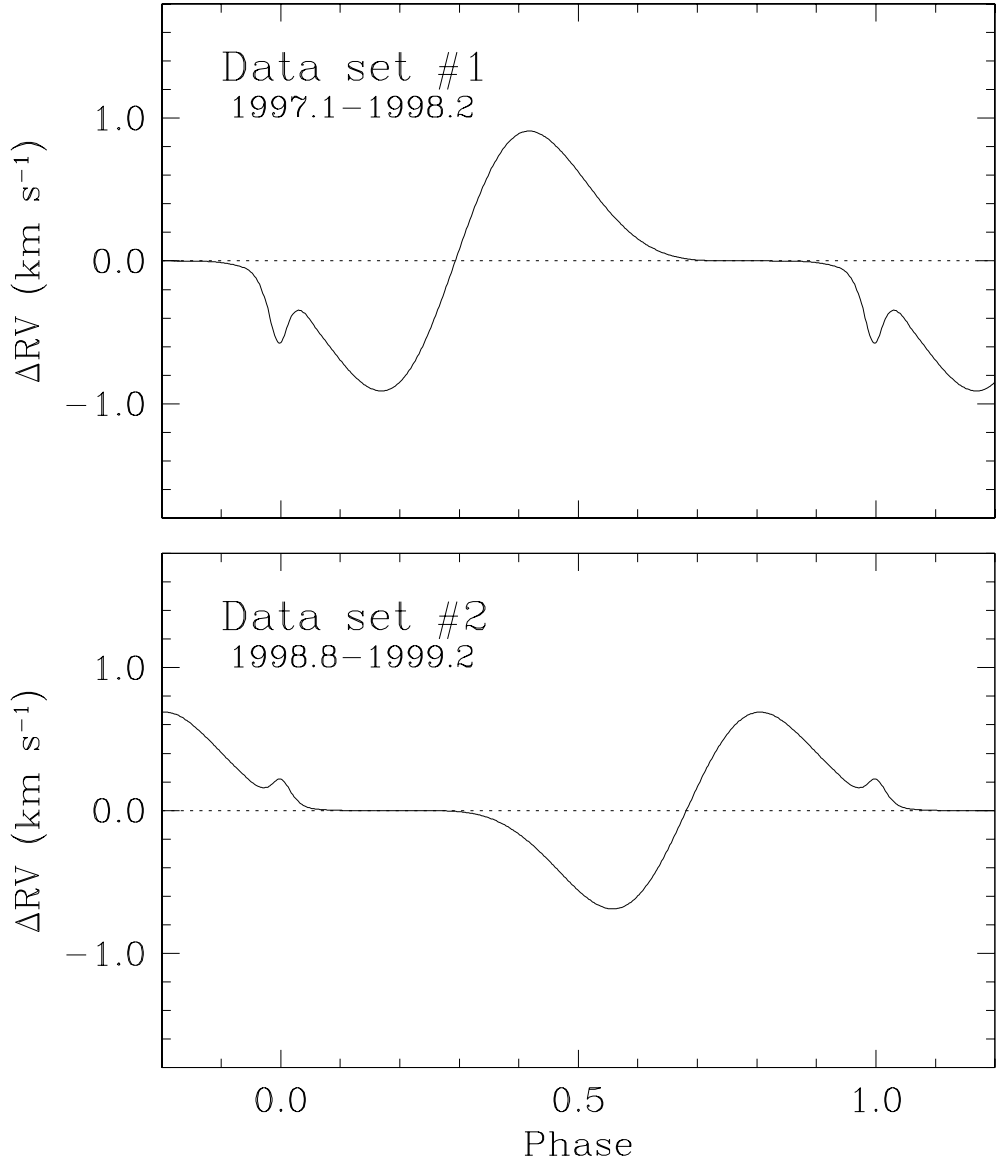}
\vskip 0.3in
 \caption[Torres.fig6.ps]{Effect of the spots on the primary of
BD$+05\arcdeg706$ upon the measured radial velocities of that
component at the mean epoch of data sets \#1 and
\#2.\label{fig:rvspot}}
 \end{figure}

While the corrections described in the preceding paragraph (which
depend only on the geometry of the system) do not change with time,
the surface features modeled in the light curves, which can also
affect the measured velocities, do change significantly as seen in
\S\ref{sec:phot}.  Spots have an effect on the measured velocities
because they too contribute to break the symmetry of the intensity
distribution on the stellar disks. In this case the effect will be on
the primary velocities because the spots are assumed to be on that
star. The phase dependence of the effect is seen in
Figure~\ref{fig:rvspot} separately for data sets \#1 and \#2, and was
derived once again using the Wilson-Devinney code. Specifically, the
effect is calculated as the difference between the velocities from
purely Keplerian motion and those computed by integration over the
visible disk of the star, with its non-uniform brightness distribution
due to the spots.  It is unclear whether the velocities we have
measured by cross-correlation are affected by the full amplitude of
the correction computed by this program, but presumably they will be
biased to some extent in the direction indicated.  Unfortunately, as
seen in Figure~\ref{fig:hist} our spectroscopic observations were not
simultaneous with either of the photometric data sets, so that
corrections for this effect cannot be applied. The magnitude of the
distortion in the velocities can reach values as large as 1~\kms\
(Figure~\ref{fig:rvspot}).  Nevertheless, as a test we applied these
corrections to our velocities separately for both data sets and
recomputed the spectroscopic orbit to explore the impact on the
elements. For data set \#1 the changes in all derived quantities
including the minimum masses are of the order of 1$\sigma$ or smaller.
For data set \#2 they are less than 0.5$\sigma$. 

Our final spectroscopic orbital solution was obtained using the
measured velocities corrected for proximity effects and for the
Rossiter distortions only.  The minimum masses and the mass ratio we
determine are only slightly different from those reported in Paper~I,
but we now find that the eccentricity is indistinguishable from zero.
The original solution in Paper~I gave a value of $e = 0.0141 \pm
0.0026$, more than 5$\sigma$ different from zero.  This presented
something of a puzzle given that tidal forces for stars with deep
convective envelopes such as those in BD$+05\arcdeg$706 are expected
to circularize the orbit very quickly for the relatively short period
in this system, considering that both components are giants. The
distortion corrections applied here appear to have solved this
dilemma, and a circular orbit is adopted hereafter. 

The revised elements of the spectroscopic orbit are given in
Table~\ref{tab:specelem}. The ephemeris for the primary eclipse that
we derive from this orbit is that given in eq.(1), and was adopted for
the light-curve solutions. The light ratio at the wavelength of our
observations (5187~\AA), computed as described by \citet{Zucker1994},
is $L_2/L_1 = 1.06 \pm 0.03$. A slight correction of $+0.06$ is
required to convert this to the visual band (see Paper~I). The
residuals of the individual observations from the orbital fit are
listed in columns 8 and 9 of Table~\ref{tab:rvs}. 
	
\section{Absolute dimensions}
\label{sec:absdim}

The light elements in the last column of Table~\ref{tab:ds1} and
Table~\ref{tab:ds2} combined with the spectroscopic elements in
Table~\ref{tab:specelem} lead to the absolute masses and radii of the
components, along with other derived properties. As mentioned earlier,
the uncertainties listed for the light elements are strictly internal
errors reported by the Wilson-Devinney code. More realistic errors for
the inclination angle, the primary and secondary potentials, and for
the secondary temperature were adopted based in part on the agreement
between the solutions in the different passbands and on the
sensitivity of the fits to the various parameters: $i = 79\fdg47 \pm
0\fdg15$, $\Omega_1 = 6.03 \pm 0.15$, $\Omega_2 = 2.246 \pm 0.011$,
and $T_2 = 4640 \pm 150$~K.  The physical properties of the stars are
summarized in Table~\ref{tab:absdim}. 

The absolute masses are determined with errors of 1.0\% and 1.7\% for
the primary and secondary, respectively, while the radii are good to
2.6\% and 1.9\%. Since the difference in the depths of the minima is a
very sensitive indicator of the temperature difference between the
components, the later quantity is typically best determined from the
light curves. In this case the result ($\Delta T_{\rm eff} = 360$~K)
is excellent agreement with our spectroscopic determination of $\Delta
T_{\rm eff} = 400$~K inferred in \S\ref{sec:spec}, based on grids of
cross-correlations with synthetic spectra. The ratio of the
luminosities in the visual band from the light curve solutions is
$(L_2/L_1)_V = 1.44 \pm 0.05$, whereas the estimate derived from the
bolometric luminosities and bolometric corrections
(Table~\ref{tab:absdim}) is $(L_2/L_1)_V = 1.29 \pm 0.26$.  The first
of these values is larger than the spectroscopic estimate
[$(L_2/L_1)_V = 1.12 \pm 0.03$], a discrepancy possibly related to the
presence of spots.  

The measured projected rotational velocity of the secondary component,
$v_2 \sin i$, is consistent with the predicted value for synchronous
rotation within the errors, as expected from the fact that the star
is filling its critical surface.  On the other hand, the rotation of
the primary is formally some 15\% faster than the predicted value, a
2.8$\sigma$ difference that may perhaps be significant. The primaries
of many other Algol systems have also been found to rotate
supersynchronously, and in some cases even close to the centrifugal
limit \citep[see, e.g.,][]{vanHamme1990}. This spin-up is usually
explained by angular momentum transfer through accretion
\citep{Huang1966}. 

The inferred distance to BD$+05\arcdeg$706 of 595 $\pm$ 51~pc relies
on the apparent brightness of the system ($V = 9.45 \pm 0.03$), the
interstellar extinction adopted in Paper~I ($A_V = 0.35 \pm 0.15$),
and bolometric corrections from \cite{Flower1996}. 
	
\section{X-ray light curve}
\label{sec:xray}

\begin{figure}
\vskip -0.3in
\epsscale{0.9}
\plotone{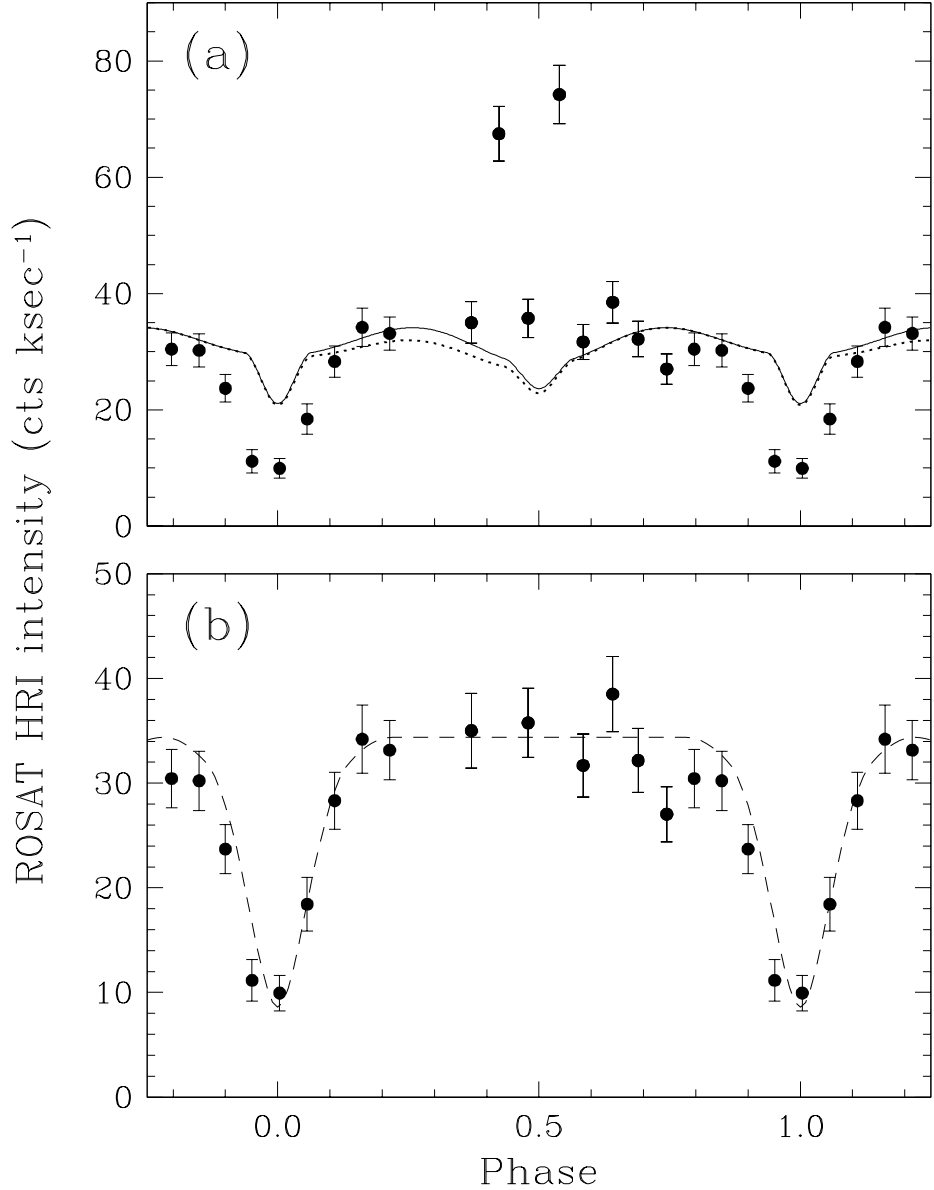}
\vskip 0.7in
 \caption[Torres.fig7.ps]{(a) ROSAT broad-band X-ray observations
of BD$+05\arcdeg706$ during a full orbital cycle, and optical light
curve superimposed for comparison ($R$ band from data set \#1). The
solid curve corresponds to the fit to the optical data with the
distortions due to the cool spot removed, while the dotted curve
includes the effect of the spot.  The normalization is carried out at
phase 0.75; (b) Toy model computed with the Wilson-Devinney code (see
text) that provides a reasonably good match to the observed depth and
duration of the primary eclipse (outliers near phase 0.5
ignored).\label{fig:xraycurve}}
 \end{figure}

The count rates over the full spectral range (0.1--2.4 keV) of the
X-ray observations of BD$+05\arcdeg$706 carried out with the ROSAT
satellite are shown as a function of orbital phase in
Figure~\ref{fig:xraycurve}a.  Two features of this broad-band light
curve are immediately obvious: (a) While the primary eclipse is
clearly present, there is no sign of a secondary eclipse near phase
0.5; (b) One measurement immediately preceding phase 0.5 and another
one following it are about a factor of two higher than the average
level of the neighboring points. 

The first of these features supports the notion that the primary is
the X-ray active star in the system, and that the secondary
contributes little, if anything, to the total X-ray luminosity of the
binary. This observation is consistent with the activity evidence
presented in Paper~I, where the primary was shown to be the dominant
source of the H$\alpha$ emission.  As reported there, several red
spectra of the binary outside of eclipse displayed relatively weak
H$\alpha$ absorption (residual intensity $\approx 0.8$), whereas one
spectrum obtained during primary eclipse displayed a much deeper
H$\alpha$ line (residual intensity $\approx 0.2$). This change was
interpreted as an indication that the emission from the active primary
fills in the normal absorption line of the secondary at most phases in
the orbit making it appear weak, except near phase 0.0, where the
primary is blocked and the secondary dominates the spectrum. 

The second feature ---unusually large X-ray flux of the two
observations near phase 0.5--- could possibly be due to two separate
flaring events, which would not be uncommon in this type of system. We
discuss this further below. 
	
A comparison with the light curves in the optical reveals two other
notable characteristics of the X-ray light curve: (c) The duration of
the primary eclipse in X-rays is much longer than in the optical; (d)
The primary minimum is considerably deeper in X-rays than it is in the
optical. To illustrate these two features we have superimposed on
Figure~\ref{fig:xraycurve}a the light curve in the $R$ band from data
set \#1, which is contemporaneous with the ROSAT observations. The
optical light curve is normalized to agree with the average level of
the X-ray curve outside of eclipse. We note that at the epoch of these
observations the configuration of the system inferred in
\S\ref{sec:phot} includes a cool spot on the primary that is on the
far side of the star at phase 0.75 and therefore does not affect the
$R$-band light curve at the second quadrature. The solid line in
Figure~\ref{fig:xraycurve}a represents the modeled $R$-band light
curve without the effect of the spot, and the dotted line includes the
distortion due to the spot, which makes the system fainter at the
first quadrature. 

It can be seen from the figure that the total duration of the primary
eclipse in the optical is approximately 0.11 in phase ($\sim$2.1~d),
while the duration in X-rays is roughly 2.9 times longer ($\sim$0.32
in phase, or $\sim$6.0~d). This implies that the size of the X-ray
emitting region around the primary must be larger than the star itself
by approximately the same factor. Since the radius of the star in
terms of the separation is $R_1/a = 0.1720$ (parameter $r_{\rm vol}$
in Table~\ref{tab:ds1}), it follows that the radius of the associated
X-ray region is $R_X/a \approx 0.50$. We note also that the mass ratio
of the binary leads independently to a mean radius for the critical
surface of the primary of $R_1^{\rm Roche}/a = 0.53$. Thus, the size
of the X-ray emitting region around the primary happens to be very
close to that of its Roche lobe ---possibly not a coincidence--- and
one may speculate that the X-ray emitting plasma is perhaps contained
within the critical surface.  Since there is no evidence from the
X-ray light curve of any significant emission from the secondary, one
might also speculate that the coronal region around that component
represents a relatively thin layer that does not extend significantly
above the photosphere of the star, which already fills its Roche lobe.
As a result, BD$+05\arcdeg$706 would appear to exhibit the curious
property of being essentially an overcontact system in X-rays, and a
semidetached system in the optical\footnote{For an illuminating
discussion of the meaning of the word ``overcontact", we refer the
reader to the recent papers by \cite{Rucinski1997} and
\cite{Wilson2001}.}. 

At optical wavelengths ($R$ band) the brightness of BD$+05\arcdeg$706
during the primary minimum drops to about 62\% of the level outside of
eclipse (phase 0.75). In X-rays the eclipse is much deeper, reaching
down to $\sim$30\% of maximum. Since blocking by the same star (the
secondary) produces significantly different depths in the optical and
in X-rays, and given that the apparent size of the primary in X-rays
is much larger, one possible explanation for the deeper minimum than
in the optical may be that the X-ray brightness distribution on the
primary is considerably more concentrated towards the center of the
disk (or toward the inner Lagrangian point, which is projected not far
from the center) than in the optical.  In order to simulate this
situation with light-curve solution programs such as Wilson-Devinney,
not originally intended for this type of problem, the relative sizes
of the stars must be changed compared to the geometry inferred from
the optical light curves, effectively by altering the mass ratio.
Furthermore, variations in the predicted light curve near phase 0.5
due to the changing aspect ratio of the primary (if it is assumed to
fill its critical surface in X-rays) must be ignored as they probably
do not apply at X-ray wavelengths.  We find a reasonably good
agreement between such a ``model" and the observed light curve when
the effective size of the X-ray emitting region of the primary is
reduced to approximately the same size as the photosphere of the
lobe-filling secondary (see Figure~\ref{fig:xraycurve}b). Of course,
any parameters derived in this way are physically meaningless and are
only intended to achieve a qualitatively good fit. The reference
provided by this very crude model does seem to suggest, nevertheless,
that the X-ray flux preceding the primary eclipse (beginning at phase
0.75, and perhaps including the descending branch) is somewhat lower
than the flux after the minimum at symmetrical phases. 

As mentioned above, the two high X-ray points near phase 0.5 might be
due to two separate flare events (since an observation intermediate
between these two shows a normal flux level). X-ray flares are common
in RS~CVn systems \citep[see, e.g.,][]{Siarkowski1996,Garcia2003}.  On
the other hand, the semidetached configuration in the optical is an
indication that mass transfer by Roche-lobe overflow may be occurring.
Due to the relatively large size of the primary, a stream of material
from the secondary to the primary would actually impact the primary
star directly rather than sweep past it and form a disk \citep[][see
also Figure~11 in Paper~I]{Lubow1975}.  Given the orbital phases at
which the discrepant X-ray points in Figure~\ref{fig:xraycurve}a occur
(0.423 and 0.539), just before and just after central blocking of the
secondary by the primary, it is possible that the sudden increase in
X-ray flux is related to the viewing angle of the stream associated
with mass transfer, which would be roughly aligned with the line of
sight at these phases. Detailed hydrodynamical modeling is required to
investigate this further. 

\begin{figure}
\vskip -0.3in
\epsscale{0.9}
\plotone{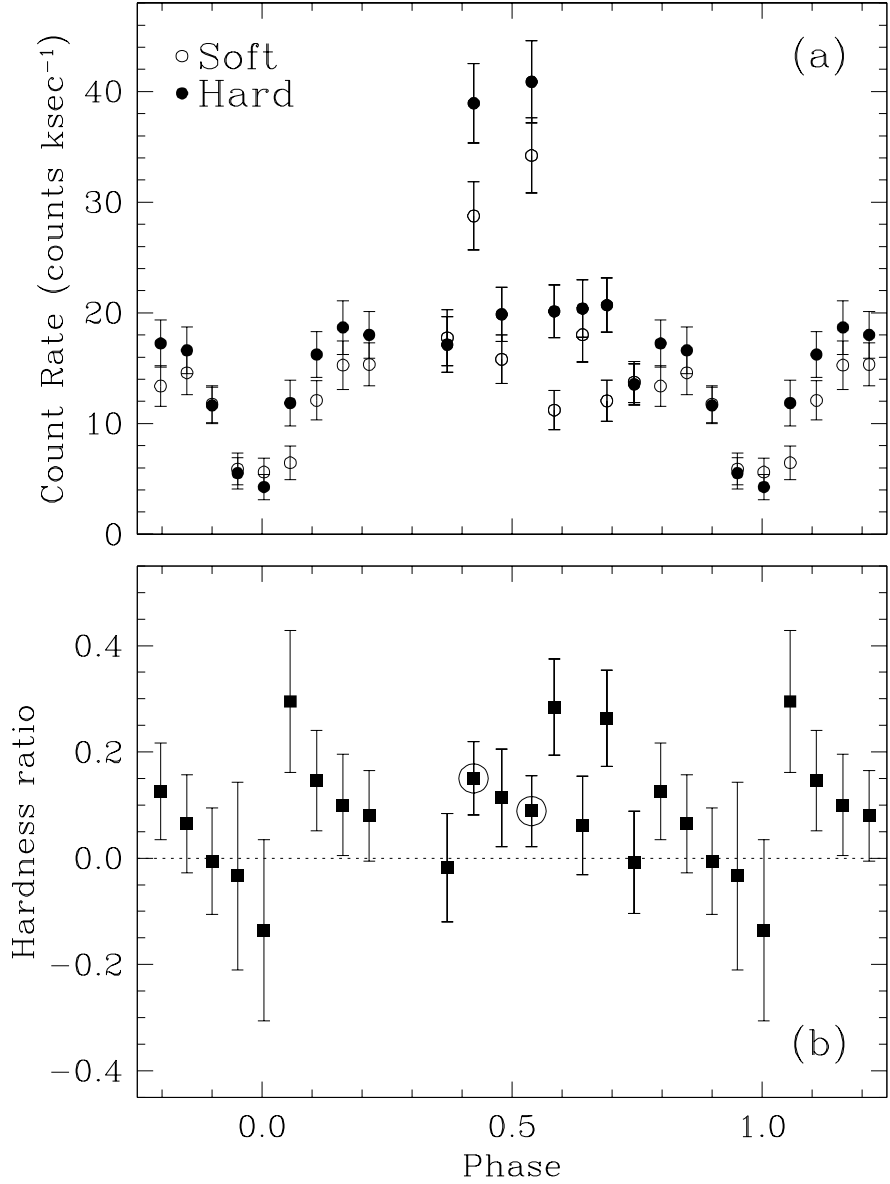}
\vskip 0.7in
 \figcaption[Torres.fig8.ps]{(a) ROSAT X-ray light curve of
BD$+05\arcdeg706$ separated into a soft band (0.1--0.5~keV) and a hard
band (0.6--2.4~keV), as a function of orbital phase. (b) Hardness
ratio (see text) for each ROSAT observation, as a function of phase.
The two observations from the broad-band light curve
(Figure~\ref{fig:xraycurve}) showing increased flux, possibly due to
flares, are marked with circles.\label{fig:hardx}}
 \end{figure}

In Figure~\ref{fig:hardx} we have split the ROSAT X-ray observations
of BD$+05\arcdeg$706 into a soft band (0.1--0.5~keV) and a hard band
(0.6--2.4~keV) in order to investigate the possibility that there
might be a difference in the behavior as a function of orbital phase.
The run of the count rates turns out to be quite similar, however,
with the soft band being slightly lower throughout
(Figure~\ref{fig:hardx}a). The changes in the hardness ratio as
defined in \S\ref{sec:obs} are shown in Figure~\ref{fig:hardx}b.
Within the errors the {\it HR\/} values are fairly constant, the
decrease toward the primary minimum being only marginally significant.
In particular, the observations near phase 0.5 pointed out in the
broad-band light curve (Figure~\ref{fig:xraycurve}a) that are possibly
associated with flaring episodes show no sign that the X-ray energy
becomes significantly harder (observations marked with circles), as
would be expected if the temperature were to increase due to heating
of the plasma. 
	
\section{Discussion and concluding remarks}
\label{sec:discussion}

The combination of our new $BV\!RI$ photometry with a reanalysis of
the spectroscopy reported in Paper~I has allowed us to obtain highly
precise absolute masses and radii for both components of the
semidetached system BD$+05\arcdeg$706, as well as the effective
temperatures and other properties. Among the dozen or so known cool
Algols, this is now by far the one with the best determined physical
parameters. Only three others have published light-curve solutions to
date (RZ~Cnc, AR~Mon, and RV~Lib). 

In Paper~I the eclipsing nature of BD$+05\arcdeg$706 was not yet
known, so there was no direct information on the orbital inclination
angle although some constraints were available based on other
evidence.  Qualitative conclusions on the history and evolutionary
state of the system were drawn there based upon the assumption that
the inclination angle was large, and therefore that the true values of
the masses and the radii were not far from the lower limits derived
for those quantities. We can now confirm the validity of that
assumption in Paper~I, and consequently also of those conclusions: the
present configuration of BD$+05\arcdeg$706 was probably reached as a
result of Case~B mass transfer \citep{Kippenhahn1967}, and the initial
mass ratio of the system is likely to have been close to unity
\citep[see][]{Eggleton1992,Nelson2001} so that the original secondary
had enough time to begin its evolution off the main sequence before
the original primary reached the point of Roche-lobe overflow and
started transferring mass onto it (eventually leading to a mass ratio
reversal).  The minimum amount of mass exchanged (under conservative
mass transfer) is $\Delta M = M_1 - (M_1+M_2)/2 = 1.05$~M$_{\sun}$.
Thus, the secondary appears to have lost at least 2/3 of its original
mass.  As noted in our earlier study of this system, although the
current mass ratio ($q \approx 0.21$) is quite typical of classical
Algol systems the relatively cool temperature of the current primary
(5000~K) is not.  This difference with the much hotter primaries of
the classical Algols is at least qualitatively well explained by the
ideas about the evolution of cool Algols mentioned above and discussed
at length in Paper~I. 

Evidence from our optical light curves suggests the presence of large
surface inhomogeneities (spots) which are quite common in similarly
active systems. Significant shifts in the location (longitude) of
these features are seen in BD$+05\arcdeg$706 over a period of 1-2
years, which are also consistent with changes seen in other RS~CVn
binaries.  Our X-ray light curve showing only the primary eclipse
confirms the report in Paper~I that the hotter and more massive
primary star is the dominant site of activity in this binary. It is
interesting to note that this is not typically the case in RS~CVn
systems, where the cooler star is usually the more magnetically active
component.  The X-ray observations also offer interesting clues about
the geometry and other properties of the X-ray emitting region, which
may be confined to the size of the critical lobe of the current
primary star. 

Photometric and spectroscopic studies of other binaries systems
identified as cool Algols are currently underway as part of a
long-term project to study this very interesting class of active
stars.  Detailed analyses of those cases along the lines of the one
presented here for BD$+05\arcdeg$706 are expected to provide further
glimpses into their properties and evolutionary histories. 
	
\acknowledgements

We thank the anonymous referee for numerous comments and helpful
suggestions.  This research has made use of the SIMBAD database,
operated at CDS, Strasbourg, France, and of NASA's Astrophysics Data
System Abstract Service. RN wishes to acknowledge financial support
from the Bundesministerium f\"ur Bildung und Forschung through the
Deutsche Zentrum f\"ur Luft- und Raumfahrt e.V.\ (DLR) under grant
number 50~OR~0003. LM would like to acknowledge the help of Julia
Lynch and Akbar Rizvi in making these observations. The Gettysburg
College Observatory is supported by Gettysburg College with additional
funding for the CCD camera provided by the National Science
Foundation.

\clearpage

\clearpage

\clearpage

\clearpage

\clearpage

\clearpage

\clearpage

\clearpage

\clearpage

\begin{deluxetable}{ccccc}
\tablenum{1}
\tablewidth{38pc}
\tablecaption{Program stars\label{tab:compcheck}}
\tablehead{
\colhead{~~~~~~~~~~~~~~~~~~~~~Star~~~~~~~~~~~~~~~~~~~~~} & 
\colhead{R.A. (J2000.0) Dec.} & 
\colhead{Sp.T.} & 
\colhead{$V$} & 
\colhead{$B\!-\!V$} 
}
\startdata
BD$+05\arcdeg$706 \dotfill & $04^{\rm h}41^{\rm m}57.6^{\rm s}$ $+05^\circ36^\prime34^{\prime\prime}$ & G5    & \phn 9.45 & 1.11 \\
TYC~95-1495-1 (comparison) \dotfill        & $04^{\rm h}41^{\rm m}36.0^{\rm s}$ $+05^\circ43^\prime10^{\prime\prime}$ &       &     10.01 & 1.19 \\
HD~29764 (check) \dotfill             & $04^{\rm h}41^{\rm m}28.7^{\rm s}$ $+05^\circ33^\prime53^{\prime\prime}$ & F8    & \phn 9.49 & 0.49 \\
\enddata
\end{deluxetable}

\clearpage

\begin{deluxetable}{c@{}c@{~~~}c@{}c@{~~~}c@{}c@{~~~}c@{}c}
\tablenum{2}
\tablewidth{34pc}
\tabletypesize{\scriptsize}
\tablecaption{Differential magnitudes in the $B$ band, in the sense
$\langle$variable$-$comparison$\rangle$.\label{tab:obsb}}
\tablehead{
\colhead{HJD} & \colhead{$\Delta B$} & \colhead{HJD} & \colhead{$\Delta B$} &
\colhead{HJD} & \colhead{$\Delta B$} & \colhead{HJD} & \colhead{$\Delta B$} \\
\colhead{(2,450,000+)} & \colhead{(mag)} & \colhead{(2,450,000+)} & \colhead{(mag)} &
\colhead{(2,450,000+)} & \colhead{(mag)} & \colhead{(2,450,000+)} & \colhead{(mag)} 
}
\startdata
  772.6936  &  $-$0.412  &   827.7321  &  $-$0.457  &  \phn903.5477  &  $-$0.486  &  1208.6844  &  $-$0.282  \\
  772.7038  &  $-$0.400  &   828.5234  &  $-$0.447  &  \phn914.5239  &  $-$0.328  &  1209.5255  &  $+$0.113  \\
  773.6906  &  $-$0.315  &   834.5772  &  $-$0.396  &  \phn915.5189  &  $-$0.233  &  1209.5601  &  $+$0.095  \\
  777.6886  &  $-$0.341  &   840.5560  &  $-$0.156  &  1133.7158  &  $+$0.017  &  1209.5601  &  $+$0.120  \\
  777.7616  &  $-$0.341  &   844.5362  &  $-$0.459  &  1136.7110  &  $-$0.408  &  1209.6537  &  $+$0.125  \\
  778.6654  &  $-$0.418  &   845.5979  &  $-$0.492  &  1140.6862  &  $-$0.421  &  1209.7217  &  $+$0.133  \\
  784.6604  &  $-$0.085  &   846.5413  &  $-$0.491  &  1142.6528  &  $-$0.223  &  1210.5415  &  $-$0.256  \\
  784.7649  &  $-$0.089  &   853.6187  &  $-$0.402  &  1148.6430  &  $-$0.434  &  1216.5883  &  $-$0.362  \\
  795.6793  &  $-$0.307  &   854.6542  &  $-$0.445  &  1149.6591  &  $-$0.411  &  1219.5979  &  $-$0.104  \\
  798.6679  &  $-$0.450  &   855.5462  &  $-$0.435  &  1157.6063  &  $-$0.492  &  1220.6234  &  $-$0.230  \\
  799.6417  &  $-$0.421  &   857.5909  &  $-$0.328  &  1162.6686  &  $-$0.069  &  1221.6098  &  $-$0.306  \\
  800.6508  &  $-$0.342  &   859.5101  &  $-$0.137  &  1165.6690  &  $-$0.371  &  1225.5947  &  $-$0.388  \\
  802.6055  &  $-$0.172  &   859.6675  &  $-$0.094  &  1166.6970  &  $-$0.440  &  1229.5403  &  $-$0.332  \\
  802.7780  &  $-$0.158  &   860.6232  &  $-$0.200  &  1170.6638  &  $-$0.285  &  1230.5336  &  $-$0.420  \\
  803.7220  &  $-$0.114  &   870.6699  &  $-$0.275  &  1173.6636  &  $-$0.382  &  1231.5202  &  $-$0.459  \\
  809.5834  &  $-$0.473  &   878.5792  &  $-$0.091  &  1180.6680  &  $-$0.187  &  1232.5722  &  $-$0.504  \\
  814.5945  &  $-$0.409  &   889.5291  &  $-$0.309  &  1190.6157  &  $+$0.114  &  1268.5297  &  $-$0.405  \\
  824.6819  &  $-$0.397  &   896.5880  &  $-$0.273  &  1198.6265  &  $-$0.303  &  1275.5187  &  $-$0.083  \\
  825.5523  &  $-$0.462  &   897.5484  &  $-$0.107  &  1207.6736  &  $-$0.346  &  1276.5165  &  $-$0.163  \\
  827.5643  &  $-$0.492  &   899.5315  &  $-$0.351  &  1208.5180  &  $-$0.303  &  1277.5134  &  $-$0.237  \\
\enddata
\end{deluxetable}
 
\clearpage

\begin{deluxetable}{c@{}c@{~~~}c@{}c@{~~~}c@{}c@{~~~}c@{}c}
\tablenum{3}
\tablewidth{34pc}
\tabletypesize{\scriptsize}
\tablecaption{Differential magnitudes in the $V$ band, in the sense 
$\langle$variable$-$comparison$\rangle$.\label{tab:obsv}}
\tablehead{
\colhead{HJD} & \colhead{$\Delta V$} & \colhead{HJD} & \colhead{$\Delta V$} & 
\colhead{HJD} & \colhead{$\Delta V$} & \colhead{HJD} & \colhead{$\Delta V$} \\ 
\colhead{(2,450,000+)} & \colhead{(mag)} & \colhead{(2,450,000+)} & \colhead{(mag)} &
\colhead{(2,450,000+)} & \colhead{(mag)} & \colhead{(2,450,000+)} & \colhead{(mag)} 
}
\startdata
  772.6907  &  $-$0.432  &   827.5608  &  $-$0.524  &\phn899.5290 &  $-$0.356  &  1208.5156  &  $-$0.331  \\
  773.6792  &  $-$0.385  &   827.7283  &  $-$0.494  &\phn903.5449 &  $-$0.500  &  1208.6819  &  $-$0.312  \\
  777.6699  &  $-$0.392  &   828.5199  &  $-$0.476  &\phn914.5212 &  $-$0.352  &  1209.5229  &  $+$0.046  \\
  777.6784  &  $-$0.393  &   834.5746  &  $-$0.423  &\phn915.5230 &  $-$0.301  &  1209.5577  &  $+$0.031  \\
  777.7580  &  $-$0.401  &   840.5508  &  $-$0.167  &  1117.7249  &  $-$0.405  &  1209.6508  &  $+$0.041  \\
  778.6590  &  $-$0.449  &   840.5621  &  $-$0.182  &  1118.7028  &  $-$0.468  &  1209.7102  &  $+$0.035  \\
  784.6547  &  $-$0.102  &   842.6729  &  $-$0.344  &  1129.6644  &  $-$0.468  &  1209.7191  &  $+$0.073  \\
  784.7536  &  $-$0.107  &   844.5336  &  $-$0.496  &  1130.7223  &  $-$0.434  &  1210.5387  &  $-$0.273  \\
  785.6335  &  $-$0.304  &   845.5953  &  $-$0.523  &  1133.7128  &  $-$0.040  &  1215.6181  &  $-$0.456  \\
  795.6681  &  $-$0.350  &   846.5346  &  $-$0.508  &  1136.7083  &  $-$0.421  &  1216.5854  &  $-$0.367  \\
  797.6278  &  $-$0.458  &   853.6160  &  $-$0.428  &  1140.6837  &  $-$0.444  &  1219.5952  &  $-$0.122  \\
  798.6644  &  $-$0.459  &   854.6436  &  $-$0.473  &  1142.6502  &  $-$0.247  &  1220.6178  &  $-$0.273  \\
  798.7559  &  $-$0.480  &   854.6458  &  $-$0.471  &  1148.6445  &  $-$0.451  &  1221.6071  &  $-$0.351  \\
  799.6356  &  $-$0.458  &   855.5436  &  $-$0.470  &  1149.6565  &  $-$0.429  &  1225.5921  &  $-$0.407  \\
  800.6471  &  $-$0.402  &   857.5882  &  $-$0.369  &  1157.6036  &  $-$0.507  &  1229.5377  &  $-$0.347  \\
  802.6019  &  $-$0.142  &   859.5074  &  $-$0.162  &  1162.6661  &  $-$0.083  &  1230.5310  &  $-$0.433  \\
  803.7185  &  $-$0.140  &   859.6649  &  $-$0.126  &  1165.6664  &  $-$0.394  &  1231.5176  &  $-$0.474  \\
  809.5799  &  $-$0.491  &   860.6207  &  $-$0.193  &  1166.6945  &  $-$0.464  &  1232.5692  &  $-$0.529  \\
  813.7879  &  $-$0.273  &   870.6672  &  $-$0.331  &  1173.6611  &  $-$0.393  &  1268.5270  &  $-$0.425  \\
  813.7911  &  $-$0.292  &   878.5685  &  $-$0.134  &  1180.6655  &  $-$0.179  &  1274.5501  &  $-$0.310  \\
  813.7931  &  $-$0.265  &   889.5291  &  $-$0.346  &  1190.6129  &  $+$0.045  &  1275.5214  &  $-$0.090  \\
  814.5910  &  $-$0.419  &   896.5854  &  $-$0.299  &  1198.6206  &  $-$0.326  &  1276.5194  &  $-$0.191  \\
  824.6780  &  $-$0.425  &   897.5454  &  $-$0.144  &  1207.6708  &  $-$0.361  &  1277.5160  &  $-$0.305  \\
  825.5623  &  $-$0.491  &             &            &             &            &             &            \\
\enddata
\end{deluxetable}
 
\clearpage

\begin{deluxetable}{c@{}c@{~~~}c@{}c@{~~~}c@{}c@{~~~}c@{}c}
\tablenum{4}
\tablewidth{34pc}
\tabletypesize{\scriptsize}
\tablecaption{Differential magnitudes in the $R$ band, in the sense 
$\langle$variable$-$comparison$\rangle$.\label{tab:obsr}}
\tablehead{
\colhead{HJD} & \colhead{$\Delta R$} & \colhead{HJD} & \colhead{$\Delta R$} & 
\colhead{HJD} & \colhead{$\Delta R$} & \colhead{HJD} & \colhead{$\Delta R$} \\ 
\colhead{(2,450,000+)} & \colhead{(mag)} & \colhead{(2,450,000+)} & \colhead{(mag)} &
\colhead{(2,450,000+)} & \colhead{(mag)} & \colhead{(2,450,000+)} & \colhead{(mag)} 
}
\startdata
  474.5480  &  $-$0.417  &   528.5478  &  $-$0.346  &\phn824.6762 &  $-$0.472  &  1142.6482  &  $-$0.327  \\
  474.5930  &  $-$0.434  &   528.5859  &  $-$0.323  &\phn825.5468 &  $-$0.538  &  1148.6430  &  $-$0.535  \\
  474.6310  &  $-$0.442  &   529.5529  &  $-$0.069  &\phn827.5594 &  $-$0.570  &  1149.6551  &  $-$0.500  \\
  474.6720  &  $-$0.436  &   529.5683  &  $-$0.083  &\phn827.7268 &  $-$0.571  &  1157.6019  &  $-$0.577  \\
  477.5058  &  $-$0.489  &   529.5919  &  $-$0.096  &\phn828.5184 &  $-$0.529  &  1162.6646  &  $-$0.147  \\
  477.5638  &  $-$0.499  &   530.5133  &  $-$0.406  &\phn834.5729 &  $-$0.481  &  1163.6604  &  $-$0.336  \\
  477.6088  &  $-$0.496  &   530.5567  &  $-$0.399  &\phn840.5600 &  $-$0.238  &  1165.6650  &  $-$0.468  \\
  477.6338  &  $-$0.497  &   530.5831  &  $-$0.414  &\phn844.5315 &  $-$0.543  &  1166.6930  &  $-$0.531  \\
  477.6518  &  $-$0.497  &   546.5490  &  $-$0.457  &\phn845.5935 &  $-$0.575  &  1170.6589  &  $-$0.399  \\
  477.6718  &  $-$0.497  &   547.5586  &  $-$0.250  &\phn846.5327 &  $-$0.573  &  1173.6598  &  $-$0.461  \\
  483.4803  &  $-$0.384  &   548.5461  &  $-$0.096  &\phn853.6143 &  $-$0.483  &  1180.6640  &  $-$0.251  \\
  483.4933  &  $-$0.396  &   549.5267  &  $-$0.397  &\phn854.6396 &  $-$0.526  &  1190.6113  &  $-$0.043  \\
  483.6473  &  $-$0.417  &   552.5314  &  $-$0.497  &\phn855.5412 &  $-$0.532  &  1198.6206  &  $-$0.397  \\
  483.6563  &  $-$0.409  &   772.6879  &  $-$0.494  &\phn857.5866 &  $-$0.433  &  1207.6685  &  $-$0.435  \\
  489.6607  &  $-$0.470  &   773.6775  &  $-$0.442  &\phn859.5056 &  $-$0.222  &  1208.6799  &  $-$0.383  \\
  489.6847  &  $-$0.469  &   777.6683  &  $-$0.454  &\phn859.6624 &  $-$0.208  &  1209.5563  &  $-$0.074  \\
  491.5536  &  $-$0.032  &   777.7565  &  $-$0.462  &\phn859.6700 &  $-$0.192  &  1209.6492  &  $-$0.059  \\
  491.5636  &  $-$0.048  &   778.6568  &  $-$0.506  &\phn860.6190 &  $-$0.251  &  1209.7085  &  $-$0.078  \\
  491.5736  &  $-$0.041  &   784.6533  &  $-$0.159  &\phn870.6653 &  $-$0.421  &  1209.7176  &  $-$0.061  \\
  491.6055  &  $-$0.027  &   784.7522  &  $-$0.170  &\phn878.5668 &  $-$0.197  &  1210.5372  &  $-$0.346  \\
  497.6100  &  $-$0.454  &   785.6314  &  $-$0.364  &\phn889.5257 &  $-$0.408  &  1215.6097  &  $-$0.538  \\
  497.6280  &  $-$0.462  &   795.6657  &  $-$0.392  &\phn896.5838 &  $-$0.349  &  1216.5839  &  $-$0.454  \\
  497.6800  &  $-$0.461  &   797.6207  &  $-$0.508  &\phn897.5429 &  $-$0.190  &  1219.5937  &  $-$0.195  \\
  504.5271  &  $-$0.544  &   798.6629  &  $-$0.532  &\phn899.5276 &  $-$0.417  &  1220.6126  &  $-$0.356  \\
  504.5450  &  $-$0.549  &   799.6332  &  $-$0.525  &\phn903.5432 &  $-$0.561  &  1221.6048  &  $-$0.405  \\
  504.5686  &  $-$0.553  &   800.6455  &  $-$0.460  &  1117.7233  &  $-$0.473  &  1225.5905  &  $-$0.474  \\
  505.5372  &  $-$0.572  &   802.6005  &  $-$0.288  &  1118.7001  &  $-$0.522  &  1229.5362  &  $-$0.406  \\
  505.5740  &  $-$0.586  &   802.7744  &  $-$0.241  &  1118.7126  &  $-$0.524  &  1232.5674  &  $-$0.580  \\
  507.4900  &  $-$0.525  &   803.7233  &  $-$0.201  &  1129.6626  &  $-$0.527  &  1268.5255  &  $-$0.505  \\
  507.5100  &  $-$0.524  &   809.5782  &  $-$0.545  &  1130.7208  &  $-$0.508  &  1274.5253  &  $-$0.390  \\
  507.6310  &  $-$0.523  &   809.7652  &  $-$0.545  &  1133.7111  &  $-$0.136  &  1275.5227  &  $-$0.161  \\
  508.5222  &  $-$0.474  &   813.7806  &  $-$0.365  &  1136.7057  &  $-$0.494  &  1276.5209  &  $-$0.254  \\
  514.5140  &  $-$0.494  &   814.5891  &  $-$0.458  &  1140.6819  &  $-$0.506  &  1277.5194  &  $-$0.384  \\
  514.6029  &  $-$0.499  &             &            &             &            &             &            \\
\enddata
\end{deluxetable}
 
\clearpage

\begin{deluxetable}{c@{}c@{~~~}c@{}c@{~~~}c@{}c@{~~~}c@{}c}
\tablenum{5}
\tablewidth{34pc}
\tabletypesize{\scriptsize}
\tablecaption{Differential magnitudes in the $I$ band, in the sense 
$\langle$variable$-$comparison$\rangle$.\label{tab:obsi}}
\tablehead{
\colhead{HJD} & \colhead{$\Delta I$} & \colhead{HJD} & \colhead{$\Delta I$} &
\colhead{HJD} & \colhead{$\Delta I$} & \colhead{HJD} & \colhead{$\Delta I$} \\
\colhead{(2,450,000+)} & \colhead{(mag)} & \colhead{(2,450,000+)} & \colhead{(mag)} &
\colhead{(2,450,000+)} & \colhead{(mag)} & \colhead{(2,450,000+)} & \colhead{(mag)} 
}
\startdata
  474.5550  &  $-$0.489  &   514.5256  &  $-$0.575  &\phn814.5870 &  $-$0.524  &  1140.6806  &  $-$0.565  \\
  474.5950  &  $-$0.501  &   514.6011  &  $-$0.561  &\phn824.6744 &  $-$0.544  &  1142.6471  &  $-$0.389  \\
  474.6380  &  $-$0.497  &   528.5456  &  $-$0.425  &\phn825.5444 &  $-$0.594  &  1148.6419  &  $-$0.586  \\
  474.6740  &  $-$0.497  &   528.5838  &  $-$0.385  &\phn827.5575 &  $-$0.625  &  1149.6539  &  $-$0.556  \\
  477.5108  &  $-$0.552  &   529.5488  &  $-$0.169  &\phn827.7251 &  $-$0.624  &  1157.6004  &  $-$0.626  \\
  477.5908  &  $-$0.571  &   529.5664  &  $-$0.165  &\phn828.5168 &  $-$0.599  &  1162.6634  &  $-$0.214  \\
  477.6108  &  $-$0.569  &   529.5902  &  $-$0.184  &\phn834.5717 &  $-$0.547  &  1163.6587  &  $-$0.416  \\
  477.6368  &  $-$0.571  &   530.5581  &  $-$0.465  &\phn840.5587 &  $-$0.315  &  1166.6909  &  $-$0.577  \\
  477.6558  &  $-$0.583  &   530.5850  &  $-$0.491  &\phn842.6694 &  $-$0.480  &  1170.6574  &  $-$0.472  \\
  477.6758  &  $-$0.567  &   546.5476  &  $-$0.545  &\phn844.5297 &  $-$0.603  &  1173.6584  &  $-$0.527  \\
  483.4833  &  $-$0.474  &   548.5441  &  $-$0.180  &\phn845.5922 &  $-$0.636  &  1180.6629  &  $-$0.333  \\
  483.4903  &  $-$0.476  &   549.5287  &  $-$0.476  &\phn846.5297 &  $-$0.614  &  1190.6102  &  $-$0.142  \\
  483.6453  &  $-$0.481  &   552.5296  &  $-$0.541  &\phn853.6131 &  $-$0.545  &  1190.6541  &  $-$0.135  \\
  483.6603  &  $-$0.478  &   772.6852  &  $-$0.554  &\phn854.6380 &  $-$0.578  &  1198.9575  &  $-$0.466  \\
  489.6417  &  $-$0.528  &   773.6760  &  $-$0.510  &\phn855.5386 &  $-$0.569  &  1207.6672  &  $-$0.504  \\
  489.6827  &  $-$0.534  &   777.6661  &  $-$0.523  &\phn857.5851 &  $-$0.493  &  1208.5128  &  $-$0.469  \\
  491.5516  &  $-$0.132  &   777.7550  &  $-$0.533  &\phn859.5043 &  $-$0.299  &  1208.6774  &  $-$0.449  \\
  491.5616  &  $-$0.129  &   778.6511  &  $-$0.573  &\phn859.6609 &  $-$0.269  &  1208.6787  &  $-$0.466  \\
  491.5716  &  $-$0.145  &   784.6513  &  $-$0.229  &\phn860.6173 &  $-$0.307  &  1209.5551  &  $-$0.156  \\
  491.6105  &  $-$0.137  &   784.7507  &  $-$0.236  &\phn870.6636 &  $-$0.450  &  1209.6478  &  $-$0.150  \\
  497.6050  &  $-$0.532  &   785.6289  &  $-$0.447  &\phn878.5655 &  $-$0.255  &  1209.7063  &  $-$0.130  \\
  497.6250  &  $-$0.541  &   795.6635  &  $-$0.485  &\phn889.5226 &  $-$0.454  &  1209.7164  &  $-$0.145  \\
  497.6740  &  $-$0.551  &   797.6187  &  $-$0.565  &\phn896.5817 &  $-$0.432  &  1210.5362  &  $-$0.433  \\
  498.6169  &  $-$0.476  &   798.6614  &  $-$0.597  &\phn897.5411 &  $-$0.238  &  1215.6081  &  $-$0.590  \\
  498.6339  &  $-$0.483  &   799.6308  &  $-$0.579  &\phn899.5256 &  $-$0.475  &  1219.5845  &  $-$0.261  \\
  498.6569  &  $-$0.467  &   799.6353  &  $-$0.588  &\phn903.5375 &  $-$0.575  &  1220.6091  &  $-$0.416  \\
  504.5253  &  $-$0.611  &   800.6440  &  $-$0.519  &  1117.7217  &  $-$0.525  &  1221.6032  &  $-$0.496  \\
  504.5426  &  $-$0.605  &   802.5989  &  $-$0.367  &  1118.6984  &  $-$0.590  &  1225.5884  &  $-$0.521  \\
  504.5669  &  $-$0.615  &   802.7705  &  $-$0.332  &  1118.7112  &  $-$0.574  &  1229.5347  &  $-$0.466  \\
  505.5347  &  $-$0.643  &   803.7137  &  $-$0.261  &  1129.6608  &  $-$0.583  &  1232.5661  &  $-$0.615  \\
  505.5722  &  $-$0.649  &   803.7275  &  $-$0.278  &  1129.6707  &  $-$0.583  &  1268.5243  &  $-$0.530  \\
  507.4912  &  $-$0.578  &   809.5767  &  $-$0.604  &  1130.7195  &  $-$0.550  &  1275.5238  &  $-$0.227  \\
  507.5083  &  $-$0.573  &   809.7627  &  $-$0.622  &  1133.7098  &  $-$0.222  &  1276.5225  &  $-$0.319  \\
  507.6290  &  $-$0.607  &   813.7789  &  $-$0.410  &  1136.7032  &  $-$0.538  &  1277.5205  &  $-$0.456  \\
  508.5199  &  $-$0.524  &             &            &             &            &             &            \\
\enddata
\end{deluxetable}
 
\clearpage

\begin{deluxetable}{ccrrrcrrrcrrrcccc}
\rotate
\tablewidth{49pc}
\tablenum{6}
\tabletypesize{\scriptsize}
\tablecaption{ROSAT X-ray observations of BD$+05\arcdeg$706.\label{tab:xray}}
\tablehead{
\colhead{HJD} & \colhead{Exposure} & 
\multicolumn{3}{c}{Soft band (0.1--0.5 keV)} && 
\multicolumn{3}{c}{Hard band (0.6--2.4 keV)} &&
\multicolumn{3}{c}{Broad band (0.1--2.4 keV)} &&& \colhead{} \\
\cline{3-5} \cline{7-9} \cline{11-13} 
\colhead{~~~(2,400,000+)~~~} & \colhead{(s)} & 
\colhead{Counts} & \colhead{Error} & \colhead{$ML$\tablenotemark{a}} &&
\colhead{Counts} & \colhead{Error} & \colhead{$ML$\tablenotemark{a}} &&
\colhead{Counts} & \colhead{Error} & \colhead{$ML$\tablenotemark{a}} && \colhead{$HR$\tablenotemark{b}} & \colhead {$\sigma_{\rm HR}$} & \colhead{Phase}
}
\startdata
 50678.6448\dotfill &  4456 &   52.43 &  7.42 &  165.2 &&  51.85 &  7.38 &  143.7 &&  104.14 &  10.47 &  299.2 &&  $-$0.01 &  0.10 & 0.900 \\
 50679.6073\dotfill &  2940 &   17.34 &  4.24 &   61.8 &&  16.22 &  4.15 &   43.2 &&   32.98 &   5.90 &  101.3 &&  $-$0.03 &  0.18 & 0.951 \\
 50680.6024\dotfill &  3621 &   20.34 &  4.57 &   64.3 &&  15.48 &  4.10 &   32.2 &&   35.65 &   6.13 &   91.4 &&  $-$0.14 &  0.17 & 0.004 \\
 50681.6007\dotfill &  2871 &   18.55 &  4.36 &   62.0 &&  34.06 &  5.95 &  108.1 &&   52.60 &   7.38 &  169.8 &&  $+$0.29 &  0.13 & 0.056 \\
 50682.5965\dotfill &  3969 &   48.01 &  7.03 &  176.5 &&  64.45 &  8.16 &  195.2 &&  112.64 &  10.78 &  370.0 &&  $+$0.15 &  0.09 & 0.109 \\
 50683.5920\dotfill &  3262 &   49.84 &  7.11 &  224.7 &&  60.94 &  7.89 &  227.6 &&  111.23 &  10.64 &  443.0 &&  $+$0.10 &  0.10 & 0.162 \\
 50684.5872\dotfill &  4190 &   64.31 &  8.10 &  254.5 &&  75.47 &  8.79 &  260.0 &&  139.26 &  11.94 &  508.8 &&  $+$0.08 &  0.09 & 0.214 \\
 50687.5392\dotfill &  2783 &   49.43 &  7.07 &  196.7 &&  47.69 &  6.97 &  161.8 &&   97.25 &   9.94 &  349.3 &&  $-$0.02 &  0.10 & 0.371 \\
 50688.5344\tablenotemark{c}\dotfill &  3066 &   88.18 &  9.43 &  434.4 && 119.38 & 10.99 &  497.5 &&  207.54 &  14.49 &  928.3 &&  $+$0.15 &  0.07 & 0.423 \\
 50689.5924\dotfill &  3351 &   52.98 &  7.33 &  232.9 &&  66.57 &  8.24 &  242.0 &&  119.40 &  11.02 &  470.5 &&  $+$0.11 &  0.09 & 0.479 \\
 50690.7236\tablenotemark{c}\dotfill &  2993 &  102.41 & 10.15 &  490.3 && 122.37 & 11.12 &  530.9 &&  224.66 &  15.06 & 1015.0 &&  $+$0.09 &  0.07 & 0.539 \\
 50691.5826\dotfill &  3615 &   40.56 &  6.41 &  173.4 &&  72.82 &  8.64 &  237.2 &&  113.82 &  10.81 &  409.3 &&  $+$0.28 &  0.09 & 0.585 \\
 50692.6469\dotfill &  3087 &   55.64 &  7.53 &  220.6 &&  62.93 &  8.03 &  212.1 &&  118.36 &  11.01 &  427.8 &&  $+$0.06 &  0.09 & 0.641 \\
 50693.5719\dotfill &  3521 &   42.45 &  6.56 &  169.2 &&  72.87 &  8.62 &  241.5 &&  114.89 &  10.82 &  407.0 &&  $+$0.26 &  0.09 & 0.690 \\
 50694.6045\dotfill &  4054 &   55.74 &  7.54 &  164.9 &&  54.91 &  7.53 &  175.6 &&  110.07 &  10.65 &  333.2 &&  $-$0.01 &  0.10 & 0.745 \\
 50695.5993\dotfill &  4000 &   53.56 &  7.39 &  187.0 &&  68.96 &  8.47 &  199.5 &&  123.10 &  11.26 &  379.6 &&  $+$0.13 &  0.09 & 0.797 \\
 50696.5934\dotfill &  3850 &   56.17 &  7.58 &  187.5 &&  63.95 &  8.11 &  187.0 &&  119.43 &  11.07 &  373.3 &&  $+$0.06 &  0.09 & 0.850 \\
\enddata
\tablenotetext{a}{Maximum Likelihood estimator (see text).}
\tablenotetext{b}{Hardness ratio (see text).}
\tablenotetext{c}{Possible flare.}
\end{deluxetable}

\clearpage

\begin{deluxetable}{cccc}
\tablenum{7}
\tablewidth{34pc}
\tabletypesize{\scriptsize}
\tablecaption{Parameters of the light curve solutions for data set \#1\label{tab:ds1}}
\tablehead{
\colhead{~~~~~~~~~~~~~~~Parameter~~~~~~~~~~~~~~~} &
\colhead{$R$ Solution} &
\colhead{$I$ Solution} &
\colhead{$RI$ Solution} 
}
\startdata
\multicolumn{4}{l}{Geometric and radiative parameters}  \\
~~~~$i$ ($\arcdeg$)\tablenotemark{a} \dotfill & 79.55 $\pm$ 0.18\phn    
                                       & 79.46 $\pm$ 0.15\phn
                                       & 79.470 $\pm$ 0.074\phn   \\
~~~~$q\equiv M_2/M_1$          \dotfill & 0.2055
                                       & 0.2055
                                       & 0.2055          \\
~~~~$\Omega_1$\tablenotemark{a}  \dotfill & 6.05 $\pm$ 0.25 
                                       & 5.89 $\pm$ 0.16 
                                       & 6.03 $\pm$ 0.10 \\
~~~~$\Omega_2$                 \dotfill & 2.246 $\pm$ 0.011       
                                       & 2.246 $\pm$ 0.011       
                                       & 2.246 $\pm$ 0.011       \\
~~~~$r_{\rm point}$ (primary, secondary) \dotfill & 0.1718~,~0.3439
                                       & 0.1767~,~0.3439
                                       & 0.1724~,~0.3439                            \\
~~~~$r_{\rm pole}$  ~(primary, secondary) \dotfill & 0.1710~,~0.2348
                                       & 0.1758~,~0.2348
                                       & 0.1716~,~0.2348                            \\
~~~~$r_{\rm side}$  ~(primary, secondary) \dotfill & 0.1715~,~0.2442
                                       & 0.1764~,~0.2442
                                       & 0.1721~,~0.2442                            \\
~~~~$r_{\rm back}$  (primary, secondary) \dotfill & 0.1718~,~0.2767
                                       & 0.1768~,~0.2767
                                       & 0.1724~,~0.2767                            \\
~~~~$r_{\rm vol}$\tablenotemark{b}~~~(primary, secondary)\dotfill & 0.1714~,~0.2513
                                       & 0.1763~,~0.2513
                                       & 0.1720~,~0.2513                            \\
~~~~$\Delta\phi$               \dotfill & $-$0.00006 $\pm$ 0.00058\phm{$-$}          
                                       & $+$0.00097 $\pm$ 0.00064\phm{$-$}          
                                       & $+$0.00040 $\pm$ 0.00040\phm{$-$}          \\
~~~~$T_1$ (K)                  \dotfill & 5000                            
                                       & 5000                            
                                       & 5000                            \\
~~~~$T_2$ (K)\tablenotemark{a}   \dotfill & 4612 $\pm$ 47\phn\phn   
                                       & 4677 $\pm$ 40\phn\phn   
                                       & 4638 $\pm$ 11\phn\phn   \\
~~~~$L_{1,R}$\tablenotemark{c} \dotfill & 0.3956 $\pm$ 0.0091      
                                       & \nodata                         
                                       & 0.3911 $\pm$ 0.0004       \\
~~~~$L_{1,I}$\tablenotemark{c} \dotfill & \nodata                         
                                       & 0.3810 $\pm$ 0.0077      
                                       & 0.3764 $\pm$ 0.0004       \\
~~~~Albedo                     \dotfill & 0.5~,~0.5
                                       & 0.5~,~0.5                              
                                       & 0.5~,~0.5                               \\
~~~~Gravity brightening        \dotfill & 0.365~,~0.365
                                       & 0.365~,~0.365                             
                                       & 0.365~,~0.365                               \\
\multicolumn{4}{l}{Spot parameters (primary star)}  \\
~~~~$l$ ($\arcdeg$)               \dotfill & 257.77 $\pm$ 0.09\phn          
                                       & 254.44 $\pm$ 0.11\phn          
                                       & 254.39 $\pm$ 0.04\phn          \\
~~~~$b$ ($\arcdeg$)               \dotfill & 45                           
                                       & 45                           
                                       & 45                           \\
~~~~$r$ ($\arcdeg$)               \dotfill & 36.33 $\pm$ 1.35              
                                       & 37.29 $\pm$ 1.63              
                                       & 37.05 $\pm$ 0.28              \\
~~~~{\it TF\/}\tablenotemark{d}      \dotfill & 0.84                    
                                       & 0.84
                                       & 0.84 \\
\multicolumn{4}{l}{Limb darkening coefficients (linear law)}  \\
~~~~$x_{1,R}$                  \dotfill & 0.589                           
                                       & \nodata                         
                                       & 0.589                           \\
~~~~$x_{1,I}$                  \dotfill & \nodata                         
                                       & 0.481                           
                                       & 0.481                           \\
~~~~$x_{1,\rm bolo}$               \dotfill & 0.527                           
                                       & 0.527                           
                                       & 0.527                           \\
~~~~$x_{2,R}$                  \dotfill & 0.639                           
                                       & \nodata                         
                                       & 0.635                           \\
~~~~$x_{2,I}$                  \dotfill & \nodata                         
                                       & 0.511                           
                                       & 0.515                           \\
~~~~$x_{2,\rm bolo}$               \dotfill & 0.528                           
                                       & 0.527                           
                                       & 0.528                           \\
\multicolumn{4}{l}{Other quantities pertaining to the fit}  \\
~~~~$\sigma_R$ (mag)           \dotfill & 0.0126                          
                                       & \nodata                         
                                       & 0.0127                          \\
~~~~$\sigma_I$ (mag)           \dotfill & \nodata                         
                                       & 0.0131                          
                                       & 0.0133                          \\
~~~~$N_{\rm obs}$              \dotfill & 92 
                                       & 95    
                                       & 187                             \\
\enddata
\tablenotetext{a}{Fixed in the combined solution (see text).}
\tablenotetext{b}{Relative radius of a sphere with the same volume as the distorted star.}
\tablenotetext{c}{Fractional luminosity of the primary.}
\tablenotetext{d}{Fixed temperature factor ($T_{\rm spot}/T_{\rm star}$).}

\end{deluxetable}

\clearpage

\begin{deluxetable}{cccccc}
\rotate
\tablenum{8}
\tablewidth{43pc}
\tabletypesize{\tiny}
\scriptsize
\tablecaption{Parameters of the light curve solutions for data set \#2\label{tab:ds2}}
\tablehead{
\colhead{~~~~~~~~~~~~~~~Parameter~~~~~~~~~~~~~~~} &
\colhead{$B$ Solution} &
\colhead{$V$ Solution} &
\colhead{$R$ Solution} &
\colhead{$I$ Solution} &
\colhead{$BVRI$ Solution} 
}
\startdata
\multicolumn{6}{l}{Geometric and radiative parameters}  \\
~~~~$i$ ($\arcdeg$)\tablenotemark{a}  \dotfill & 79.24 $\pm$ 0.21\phn
                                    & 79.41 $\pm$ 0.25\phn
                                    & 79.38 $\pm$ 0.36\phn
                                    & 79.55 $\pm$ 0.30\phn
                                    & 79.470 $\pm$ 0.074\phn \\
~~~~$q\equiv M_2/M_1$                \dotfill & 0.2055
                                    & 0.2055
                                    & 0.2055
                                    & 0.2055
                                    & 0.2055  \\
~~~~$\Omega_1$\tablenotemark{a}  \dotfill & 6.39 $\pm$ 0.53 
                                    & 5.83 $\pm$ 0.34 
                                    & 5.83 $\pm$ 0.35 
                                    & 6.07 $\pm$ 0.32 
                                    & 6.03 $\pm$ 0.10  \\
~~~~$\Omega_2$                 \dotfill & 2.246 $\pm$ 0.011 
                                    & 2.246 $\pm$ 0.011 
                                    & 2.246 $\pm$ 0.011 
                                    & 2.246 $\pm$ 0.011 
                                    & 2.246 $\pm$ 0.011        \\
~~~~$r_{\rm point}$ (primary, secondary) \dotfill & 0.1623~,~0.3439
                                       & 0.1787~,~0.3439
                                       & 0.1787~,~0.3439
                                       & 0.1712~,~0.3439
                                       & 0.1724~,~0.3439                            \\
~~~~$r_{\rm pole}$ ~(primary, secondary) \dotfill & 0.1616~,~0.2348
                                       & 0.1777~,~0.2348
                                       & 0.1777~,~0.2348
                                       & 0.1704~,~0.2348
                                       & 0.1716~,~0.2348                            \\
~~~~$r_{\rm side}$ ~(primary, secondary) \dotfill & 0.1620~,~0.2442
                                       & 0.1783~,~0.2442
                                       & 0.1783~,~0.2442
                                       & 0.1710~,~0.2442
                                       & 0.1721~,~0.2442                            \\
~~~~$r_{\rm back}$  (primary, secondary) \dotfill & 0.1622~,~0.2767
                                       & 0.1786~,~0.2767
                                       & 0.1786~,~0.2767
                                       & 0.1712~,~0.2767
                                       & 0.1724~,~0.2767                            \\
~~~~$r_{\rm vol}$\tablenotemark{b}~~~(primary, secondary)\dotfill & 0.1620~,~0.2513
                                       & 0.1782~,~0.2513
                                       & 0.1782~,~0.2513
                                       & 0.1709~,~0.2513
                                       & 0.1720~,~0.2513                            \\
~~~~$\Delta\phi$               \dotfill & $-$0.0025 $\pm$ 0.0011\phm{$+$}
                                    & $-$0.00331 $\pm$ 0.00096\phm{$-$}  
                                    & $-$0.00329 $\pm$ 0.00088\phm{$-$} 
                                    & $-$0.00201 $\pm$ 0.00077\phm{$-$}  
                                    & $-$0.00272 $\pm$ 0.00044\phm{$-$}           \\
~~~~$T_1$ (K)                 \dotfill & 5000
                                    & 5000
                                    & 5000
                                    & 5000
                                    & 5000 \\
~~~~$T_2$ (K)\tablenotemark{a}   \dotfill & 4565 $\pm$ 67\phn\phn 
                                    & 4671 $\pm$ 52\phn\phn  
                                    & 4715 $\pm$ 56\phn\phn  
                                    & 4701 $\pm$ 59\phn\phn  
                                    & 4638 $\pm$ 11\phn\phn   \\
~~~~$L_{1,B}\tablenotemark{c}$ \dotfill &  0.443 $\pm$ 0.013 
                                    & \nodata
                                    & \nodata
                                    & \nodata
                                    & 0.4353 $\pm$ 0.0007        \\
~~~~$L_{1,V}\tablenotemark{c}$ \dotfill & \nodata
                                    & 0.418 $\pm$ 0.016 
                                    & \nodata
                                    & \nodata
                                    & 0.4102 $\pm$ 0.0006        \\
~~~~$L_{1,R}\tablenotemark{c}$ \dotfill & \nodata
                                    & \nodata
                                    & 0.390 $\pm$ 0.020 
                                    & \nodata
                                    & 0.3911 $\pm$ 0.0006        \\
~~~~$L_{1,I}\tablenotemark{c}$ \dotfill & \nodata
                                    & \nodata
                                    & \nodata
                                    & 0.361 $\pm$ 0.015 
                                    & 0.3764 $\pm$ 0.0007        \\
~~~~Albedo                     \dotfill & 0.5~,~0.5
                                       & 0.5~,~0.5
                                       & 0.5~,~0.5
                                       & 0.5~,~0.5
                                       & 0.5~,~0.5                               \\
~~~~Gravity brightening        \dotfill & 0.365~,~0.365
                                       & 0.365~,~0.365
                                       & 0.365~,~0.365
                                       & 0.365~,~0.365
                                       & 0.365~,~0.365                               \\
\multicolumn{6}{l}{Spot parameters (primary star)}  \\
~~~~$l$ ($\arcdeg$)             \dotfill & 110.29 $\pm$ 0.17\phn
                                    & 113.14 $\pm$ 0.14\phn
                                    & 117.81 $\pm$ 0.14\phn
                                    & 112.06 $\pm$ 0.14\phn
                                    & 112.73 $\pm$ 0.07\phn    \\
~~~~$b$ ($\arcdeg$)           \dotfill & 45
                                    & 45
                                    & 45
                                    & 45
                                    & 45                           \\
~~~~$r$ ($\arcdeg$)             \dotfill & 35.0  $\pm$ 1.2 
                                    & 33.0  $\pm$ 2.0 
                                    & 33.4 $\pm$ 2.2
                                    & 34.7 $\pm$ 1.6
                                    & 33.83 $\pm$ 0.65              \\
~~~~{\it TF}\tablenotemark{d}   \dotfill & 0.87 
                                    & 0.87
                                    & 0.87
                                    & 0.87
                                    & 0.87 \\
\multicolumn{6}{l}{Limb darkening coefficients (linear law)}  \\
~~~~$x_{1,B}$                  \dotfill & 0.867
                                    & \nodata
                                    & \nodata
                                    & \nodata
                                    & 0.867                           \\
~~~~$x_{1,V}$                  \dotfill & \nodata
                                    & 0.714
                                    & \nodata
                                    & \nodata
                                    & 0.714                           \\
~~~~$x_{1,R}$                  \dotfill & \nodata
                                    & \nodata
                                    & 0.589
                                    & \nodata
                                    & 0.589                           \\
~~~~$x_{1,I}$                  \dotfill & \nodata
                                    & \nodata
                                    & \nodata
                                    & 0.481
                                    & 0.481                           \\
~~~~$x_{1,\rm bolo}$               \dotfill & 0.527
                                    & 0.527
                                    & 0.527
                                    & 0.527
                                    & 0.527                           \\
~~~~$x_{2,B}$                  \dotfill & 0.944
                                    & \nodata
                                    & \nodata
                                    & \nodata
                                    & 0.931                           \\
~~~~$x_{2,V}$                  \dotfill & \nodata
                                    & 0.768
                                    & \nodata
                                    & \nodata
                                    & 0.774                           \\
~~~~$x_{2,R}$                  \dotfill & \nodata
                                    & \nodata
                                    & 0.624
                                    & \nodata
                                    & 0.635                           \\
~~~~$x_{2,I}$                  \dotfill & \nodata
                                    & \nodata
                                    & \nodata
                                    & 0.508
                                    & 0.515                           \\
~~~~$x_{2,\rm bolo}$               \dotfill & 0.529
                                    & 0.527
                                    & 0.527
                                    & 0.527
                                    & 0.528                           \\
\multicolumn{6}{l}{Other quantities pertaining to the fit}  \\
~~~~$\sigma_B$ (mag)           \dotfill & 0.0169
                                    & \nodata
                                    & \nodata
                                    & \nodata
                                    & 0.0180                          \\
~~~~$\sigma_V$ (mag)           \dotfill & \nodata
                                    & 0.0167
                                    & \nodata
                                    & \nodata
                                    & 0.0171                          \\
~~~~$\sigma_R$ (mag)           \dotfill & \nodata
                                    & \nodata
                                    & 0.0134
                                    & \nodata
                                    & 0.0145                          \\
~~~~$\sigma_I$ (mag)           \dotfill & \nodata
                                    & \nodata
                                    & \nodata
                                    & 0.0104
                                    & 0.0112                          \\
~~~~$N_{\rm obs}$              \dotfill & 37
                                       & 42
                                       & 41
                                       & 42
                                       & 162                             \\
\enddata
\tablenotetext{a}{Fixed in the combined solution (see text).}
\tablenotetext{b}{Relative radius of a sphere with the same volume as the distorted star.}
\tablenotetext{c}{Fractional luminosity of the primary.}
\tablenotetext{d}{Fixed temperature factor ($T_{\rm spot}/T_{\rm star}$).}
\end{deluxetable}

\clearpage

\begin{deluxetable}{cc@{}cc@{}cc@{}cc@{}cc}
\tablewidth{37pc}
\tablenum{9}
\tabletypesize{\tiny}
\tablecaption{Revised radial velocity measurements of BD$+05\arcdeg$706,
and residuals from the spectroscopic orbit.\label{tab:rvs}}
\tablehead{
\colhead{HJD} & \colhead{$RV_1^{\rm obs}$~~\tablenotemark{a}} & \colhead{$RV_2^{\rm obs}$~~\tablenotemark{a}} &
\colhead{$\Delta RV_1$~\tablenotemark{b}} & \colhead{$\Delta RV_2$~\tablenotemark{b}} &
\colhead{$RV_1$~\tablenotemark{c}} & \colhead{$RV_2$~\tablenotemark{c}} &
\colhead{(O-C)$_1$} & \colhead{(O-C)$_2$} & \colhead{Orbital} \\
\colhead{\hbox{~(2,400,000+)~}} & \colhead{(\kms)} & \colhead{(\kms)} &
\colhead{(\kms)} & \colhead{(\kms)} &
\colhead{(\kms)} & \colhead{(\kms)} &
\colhead{(\kms)} & \colhead{(\kms)} & \colhead{phase} \\
\colhead{(1)} & \colhead{(2)} & \colhead{(3)} & \colhead{(4)} & 
\colhead{(5)} & \colhead{(6)} & \colhead{(7)} & \colhead{(8)} & 
\colhead{(9)} & \colhead{(10)} 
}
\startdata
  49437.5420\dotfill &      $-$33.96 &    $+$79.12 &       $-$0.04 &  $+$1.16 &     $-$34.00 &    $+$80.28 &      $+$2.91 &  $+$2.70 &   0.229 \\
  49614.0033\dotfill &      $-$10.38 &    $-$55.19 & \phm{$-$}0.00 &  $+$1.44 &     $-$10.38 &    $-$53.75 &      $-$0.95 &  $+$2.34 &   0.567 \\
  49644.9114\dotfill &      $-$37.77 &    $+$72.73 &       $-$0.05 &  $+$1.06 &     $-$37.82 &    $+$73.79 &      $-$1.64 &  $-$0.24 &   0.202 \\
  49648.9088\dotfill &      $-$27.64 &    $+$35.10 &       $-$0.00 &  $-$1.50 &     $-$27.64 &    $+$33.60 &      $-$0.04 &  $+$1.27 &   0.413 \\
  49653.8368\dotfill &   \phn$+$1.52 &   $-$101.99\phn &   $+$0.02 &  $-$0.03 &  \phn$+$1.54 &   $-$102.02\phn &  $+$1.44 &  $+$0.43 &   0.674 \\
  49680.9384\dotfill &      $-$29.45 &    $+$43.32 &       $-$0.05 &  $+$0.30 &     $-$29.50 &    $+$43.62 &      $+$0.24 &  $+$0.89 &   0.108 \\
  49681.7665\dotfill &      $-$33.04 &    $+$60.85 &       $-$0.05 &  $+$0.62 &     $-$33.09 &    $+$61.47 &      $+$0.36 &  $+$0.73 &   0.152 \\
  49696.7771\dotfill &   \phn$-$9.42 &    $-$46.10 &       $-$0.76 &  $-$0.11 &     $-$10.18 &    $-$46.21 &      $+$0.66 &  $+$3.04 &   0.946 \\
  49700.7398\dotfill &      $-$33.59 &    $+$59.35 &       $-$0.05 &  $+$0.67 &     $-$33.64 &    $+$60.02 &      $+$0.08 &  $-$2.07 &   0.156 \\
  49704.7351\dotfill &      $-$34.58 &    $+$53.31 &       $-$0.01 &  $-$0.94 &     $-$34.59 &    $+$52.37 &      $-$2.61 &  $-$1.25 &   0.367 \\
  49708.6896\dotfill &   \phn$-$8.72 &    $-$63.23 & \phm{$-$}0.00 &  $+$1.49 &  \phn$-$8.72 &    $-$61.74 &      $-$0.42 &  $-$0.12 &   0.576 \\
  49732.7773\dotfill &   \phn$-$1.03 &    $-$93.39 &       $+$0.05 &  $-$0.59 &  \phn$-$0.98 &    $-$93.98 &      $+$0.54 &  $+$0.58 &   0.851 \\
  49734.8029\dotfill &   \phn$-$9.50 &    $-$40.34 &       $-$2.56 &  $-$0.08 &     $-$12.06 &    $-$40.42 &      $+$0.20 &  $+$1.88 &   0.958 \\
  49766.6378\dotfill &   \phn$-$1.68 &    $-$92.32 &       $+$0.01 &  $+$0.71 &  \phn$-$1.67 &    $-$91.61 &      $+$0.34 &  $+$0.59 &   0.643 \\
  49767.6563\dotfill &   \phn$+$0.53 &   $-$109.05\phn &   $+$0.02 &  $-$0.49 &  \phn$+$0.55 &   $-$109.54\phn &  $-$0.65 &  $-$1.71 &   0.697 \\
  49815.6120\dotfill &      $-$37.30 &    $+$78.13 &       $-$0.04 &  $+$1.15 &     $-$37.34 &    $+$79.28 &      $-$0.36 &  $+$1.35 &   0.234 \\
  49964.0085\dotfill &      $-$27.93 &    $+$30.63 &       $-$0.04 &  $+$0.20 &     $-$27.97 &    $+$30.83 &      $-$0.43 &  $-$1.17 &   0.086 \\
  49969.9892\dotfill &      $-$29.01 &    $+$39.59 & \phm{$-$}0.00 &  $-$1.44 &     $-$29.01 &    $+$38.15 &      $-$0.32 &  $+$0.53 &   0.403 \\
  49971.0119\dotfill &      $-$24.69 & \phn$+$8.75 & \phm{$-$}0.00 &  $-$2.71 &     $-$24.69 & \phn$+$6.04 &      $-$2.02 &  $-$2.27 &   0.457 \\
  50002.9115\dotfill &      $-$35.21 &    $+$57.25 &       $-$0.05 &  $+$0.55 &     $-$35.26 &    $+$57.80 &      $-$2.34 &  $-$0.40 &   0.145 \\
  50006.9548\dotfill &      $-$32.05 &    $+$58.81 &       $-$0.01 &  $-$0.75 &     $-$32.06 &    $+$58.06 &      $+$0.61 &  $+$1.11 &   0.359 \\
  50031.9081\dotfill &   \phn$-$0.28 &   $-$104.85\phn &   $+$0.02 &  $-$0.12 &  \phn$-$0.26 &   $-$104.97\phn &  $-$0.64 &  $-$1.16 &   0.679 \\
  50033.8891\dotfill &   \phn$+$2.26 &   $-$110.16\phn &   $+$0.04 &  $-$1.13 &  \phn$+$2.30 &   $-$111.29\phn &  $+$0.44 &  $-$0.26 &   0.784 \\
  50051.7831\dotfill &   \phn$+$2.68 &   $-$110.90\phn &   $+$0.03 &  $-$0.97 &  \phn$+$2.71 &   $-$111.87\phn &  $+$0.55 &  $+$0.61 &   0.731 \\
  50060.8041\dotfill &      $-$35.47 &    $+$74.57 &       $-$0.05 &  $+$1.10 &     $-$35.52 &    $+$75.67 &      $+$0.88 &  $+$0.57 &   0.208 \\
  50061.7975\dotfill &      $-$37.73 &    $+$77.49 &       $-$0.03 &  $+$1.04 &     $-$37.76 &    $+$78.53 &      $-$0.73 &  $+$0.33 &   0.261 \\
  50062.7964\dotfill &      $-$35.10 &    $+$71.98 &       $-$0.02 &  $+$0.28 &     $-$35.12 &    $+$72.26 &      $+$0.41 &  $+$1.35 &   0.313 \\
  50064.6757\dotfill &      $-$27.02 &    $+$35.36 &       $-$0.00 &  $-$1.49 &     $-$27.02 &    $+$33.87 &      $+$0.62 &  $+$1.38 &   0.413 \\
  50083.6864\dotfill &      $-$26.41 &    $+$33.18 &       $-$0.00 &  $-$1.50 &     $-$26.41 &    $+$31.68 &      $+$0.59 &  $+$2.27 &   0.419 \\
  50085.7517\dotfill &      $-$13.29 &    $-$35.19 & \phm{$-$}0.00 &  $+$3.64 &     $-$13.29 &    $-$31.55 &      $+$0.64 &  $+$2.64 &   0.528 \\
  50088.7862\dotfill &   \phn$+$0.62 &   $-$105.51\phn &   $+$0.02 &  $-$0.32 &  \phn$+$0.64 &   $-$105.83\phn &  $-$0.22 &  $+$0.32 &   0.689 \\
  50090.7706\dotfill &   \phn$+$1.38 &   $-$107.20\phn &   $+$0.05 &  $-$1.09 &  \phn$+$1.43 &   $-$108.29\phn &  $-$0.14 &  $+$1.32 &   0.794 \\
  50092.7725\dotfill &   \phn$-$5.88 &    $-$73.48 &       $+$0.05 &  $-$0.26 &  \phn$-$5.83 &    $-$73.74 &      $-$0.06 &  $+$0.17 &   0.900 \\
  50117.6706\dotfill &      $-$36.09 &    $+$75.44 &       $-$0.04 &  $+$1.13 &     $-$36.13 &    $+$76.57 &      $+$0.53 &  $+$0.21 &   0.217 \\
  50141.6862\dotfill &      $-$19.15 &    $-$12.00 & \phm{$-$}0.00 &  $-$2.62 &     $-$19.15 &    $-$14.62 &      $-$0.25 &  $-$4.60 &   0.488 \\
  50143.6293\dotfill &   \phn$-$7.21 &    $-$70.78 &       $-$0.00 &  $+$1.48 &  \phn$-$7.21 &    $-$69.30 &      $-$0.43 &  $-$0.29 &   0.591 \\
  50144.6373\dotfill &   \phn$-$3.42 &    $-$93.00 &       $+$0.01 &  $+$0.70 &  \phn$-$3.41 &    $-$92.30 &      $-$1.50 &  $+$0.38 &   0.644 \\
  50145.6448\dotfill &   \phn$+$1.11 &   $-$107.04\phn &   $+$0.02 &  $-$0.48 &  \phn$+$1.13 &   $-$107.52\phn &  $-$0.10 &  $+$0.45 &   0.697 \\
  50148.6641\dotfill &   \phn$-$3.71 &    $-$91.19 &       $+$0.05 &  $-$0.54 &  \phn$-$3.66 &    $-$91.73 &      $-$1.68 &  $+$0.62 &   0.857 \\
  50149.5956\dotfill &   \phn$-$6.31 &    $-$69.74 &       $+$0.04 &  $-$0.23 &  \phn$-$6.27 &    $-$69.97 &      $+$0.18 &  $+$0.63 &   0.906 \\
  50153.6415\dotfill &      $-$31.32 &    $+$45.86 &       $-$0.05 &  $+$0.36 &     $-$31.37 &    $+$46.22 &      $-$0.47 &  $-$2.13 &   0.120 \\
\enddata
\tablenotetext{a}{Measured velocities, including the corrections described in the text to account for the effect of the narrow spectral window.}
\tablenotetext{b}{Combined corrections due to proximity and to the Rossiter effect (see text), to be added to the measured velocities.}
\tablenotetext{c}{Final velocities used for the orbital solution.}
\end{deluxetable}

\clearpage

\begin{deluxetable}{lc}
\tablenum{10}
\tablewidth{21pc}
\tablecaption{Spectroscopic orbital elements of
BD+05$\arcdeg$706.\label{tab:specelem}}
\tablehead{
\colhead{\hfil~~~~~~~Parameter~~~~~~~~} & \colhead{Value}}
\startdata
$P$ (days)\dotfill                       &  18.8988~$\pm$~0.0011\phm{2}       \\
$\gamma$ (\kms)\dotfill                  & $-17.39$~$\pm$~0.14\phn\phs          \\
$K_1$ (\kms)\dotfill               &  19.69~$\pm$~0.22\phn                \\
$K_2$ (\kms)\dotfill               &  95.80~$\pm$~0.33\phn                \\
$e$\dotfill                              & 0 (fixed)                              \\
$T_{\rm max}$ (HJD)\tablenotemark{a}\dotfill       & 2,449,919.857~$\pm$~0.011\phm{222222,,}\\
\\
$a_1\sin i$ (10$^6$~km)\dotfill           &  5.117~$\pm$~0.059                   \\
$a_2\sin i$ (10$^6$~km)\dotfill           & 24.897~$\pm$~0.088\phn               \\
$a \sin i$ (R$_{\sun}$)\dotfill          & 43.12~$\pm$~0.15\phn                 \\
$M_1\sin^3 i$ (M$_{\sun}$)\dotfill &  2.502~$\pm$~0.026                   \\
$M_2\sin^3 i$ (M$_{\sun}$)\dotfill &  0.5143~$\pm$~0.0085                   \\
$q\equiv M_2/M_1$\dotfill    &  0.2055~$\pm$~0.0025                 \\
\\
$N_{\rm obs}$\dotfill                    & 41                                   \\
$\sigma_1$ (\kms)\dotfill          & 1.05                                 \\
$\sigma_2$ (\kms)\dotfill          & 1.57                                 \\
\enddata
\tablenotetext{a}{Time of maximum primary velocity.}
\end{deluxetable}

\clearpage

\begin{deluxetable}{lc@{~~~~~~~~}c}
\tablenum{11}
\tablewidth{28pc}
\tablecaption{Summary of system parameters for BD+05$^\circ$706.\label{tab:absdim}}
\tablehead{
\colhead{Parameter~~~~~~~~~~~~~~~~~~~}              & \colhead{Primary~~~~~~~}                            & \colhead{Secondary} 
}
\startdata
$P$\ (days)                                   \dotfill & \multicolumn{2}{c}{18.8988 $\pm$ 0.0011\phn}                            \\
$a$\ (R$_{\sun}$)                             \dotfill & \multicolumn{2}{c}{43.86 $\pm$ 0.15\phn}                                \\
$M$\ (M$_{\sun}$)                             \dotfill & 2.633    $\pm$ 0.028                         & 0.5412   $\pm$ 0.0093    \\
$R$\ (R$_{\sun}$)                             \dotfill & 7.55     $\pm$ 0.20                          & 11.02    $\pm$ 0.21\phn  \\
$T$\ (K)                                      \dotfill & 5000     $\pm$ 100\phn                       & 4640     $\pm$ 150\phn   \\
$\log\ g$                                     \dotfill & 3.103    $\pm$ 0.023                         & 2.087    $\pm$ 0.018     \\
$\bar \rho$\ ($10^{-3}$ g\ cm$^{-3}$)\tablenotemark{a}         \dotfill & 8.63     $\pm$ 0.68         & 0.569    $\pm$ 0.034     \\
$v_{\rm sync} \sin i$\ (km\ s$^{-1}$)\tablenotemark{b}   \dotfill & 19.9     $\pm$ 0.5\phn            & 29.0     $\pm$ 0.6\phn   \\
$v \sin i$\ (km\ s$^{-1}$)\tablenotemark{c}   \dotfill & 23       $\pm$ 1\phn                         & 31       $\pm$ 2\phn     \\
$\log L_{\rm bol}$\ (L$_\odot$)               \dotfill & 1.504    $\pm$ 0.041                         & 1.704    $\pm$ 0.059     \\
$\log L_X$\ (ergs\ s$^{-1}$)                  \dotfill & \multicolumn{2}{c}{31.96 $\pm$ 0.07\phn}                                \\
$L_2/L_1$                                     \dotfill & \multicolumn{2}{c}{1.58 $\pm$ 0.27}                                     \\
$M_{\rm bol}$\ (mag)                          \dotfill & 0.97     $\pm$ 0.10                          & 0.47     $\pm$ 0.15      \\
$M_V$\ (mag)                                  \dotfill & 1.26     $\pm$ 0.11                          & 0.98     $\pm$ 0.18      \\
$m-M$\ (mag)                                  \dotfill & \multicolumn{2}{c}{8.87 $\pm$ 0.18}                                     \\
Distance\ (pc)                                \dotfill & \multicolumn{2}{c}{595 $\pm$ 51\phn}                                    \\
\enddata
\tablenotetext{a}{Mean stellar density.}
\tablenotetext{b}{Projected rotational velocity based on the radius and the period, assuming synchronous rotation.}
\tablenotetext{c}{Spectroscopically measured rotational velocity.}
\end{deluxetable}
 
\end{document}